# Global GIS-based potential analysis and cost assessment of Power-to-X fuels in 2050

Maximilian Pfennig[1,a], Diana Böttger[a], Benedikt Häckner[a,b], David Geiger[a], Christoph Zink[a], André Bisevic[a], Lukas Jansen[a,b]

[a] Fraunhofer Institute of Energy Economics and Energy System Technology IEE, Joseph-Beuys-Str. 8, 34117 Kassel, Germany

[b] Department of Electrical Engineering and Computer Science, University of Kassel, Wilhelmshöher Allee 73, 34121 Kassel, Germany

## Abstract

Electricity-based fuels from renewable energies are regarded as a key instrument for climate protection. These Power-to-X (PtX) products are to replace fossil fuels in sectors where direct use of electricity from renewable energies is not possible. We investigate the production potential of electricity-based fuels from onshore wind energy and ground-mounted photovoltaic energy for all countries outside the European Economic Area along 14 PtX production pathways. These include hydrogen (gaseous and liquid), methane (gaseous and liquid), methanol, Fischer-Tropsch fuels and ammonia each with two different electrolysis technologies. The technical and economic potential assessment is based on data with hourly temporal and high spatial resolution. The analysis considers various criteria, including prevailing weather conditions, sustainability as well as nature conservation concerns and land use. The results show the production quantities and costs of climate-friendly fuel production under strict sustainability criteria and locate them spatially. It is shown that many regions of the world are well suited to produce PtX products. The production quantity outside Europe is up to 120,000 TWh$_{LHV}$/yr of hydrogen or 87,000 TWh$_{LHV}$/yr of electricity-based liquid fuels in the long term. We identify 97 countries with potentials, of which 38 countries possess relevant potentials of more than 100 TWh$_{LHV}$/yr. The largest suitable areas are in the United States, Australia and Argentina. The production costs vary a lot across the different regions. The lowest production costs for PtX generation are in Latin America (Chile, Argentina and Venezuela) and Mauritania with a lower limit of 42.3 €/MWh$_{LHV}$ to 46.5 €/MWh$_{LHV}$ for gaseous hydrogen and 84 €/MWh$_{LHV}$ to 89.1 €/MWh$_{LHV}$ for Fischer-Tropsch fuels. All best sites are pure wind or combined wind and photovoltaic sites. Moreover, we show import options of these PtX products to Europe considering socioeconomic factors and different transport options. All investigation results are freely accessible via the Global PtX Atlas on https://maps.iee.fraunhofer.de/ptx-atlas/.

## Keywords

Power-to-X; global analysis; cost optimisation; GIS-based potential analysis; hydrogen; E-fuels; ammonia

---

[1] Corresponding author: maximilian.pfennig@iee.fraunhofer.de

# Graphical Abstract

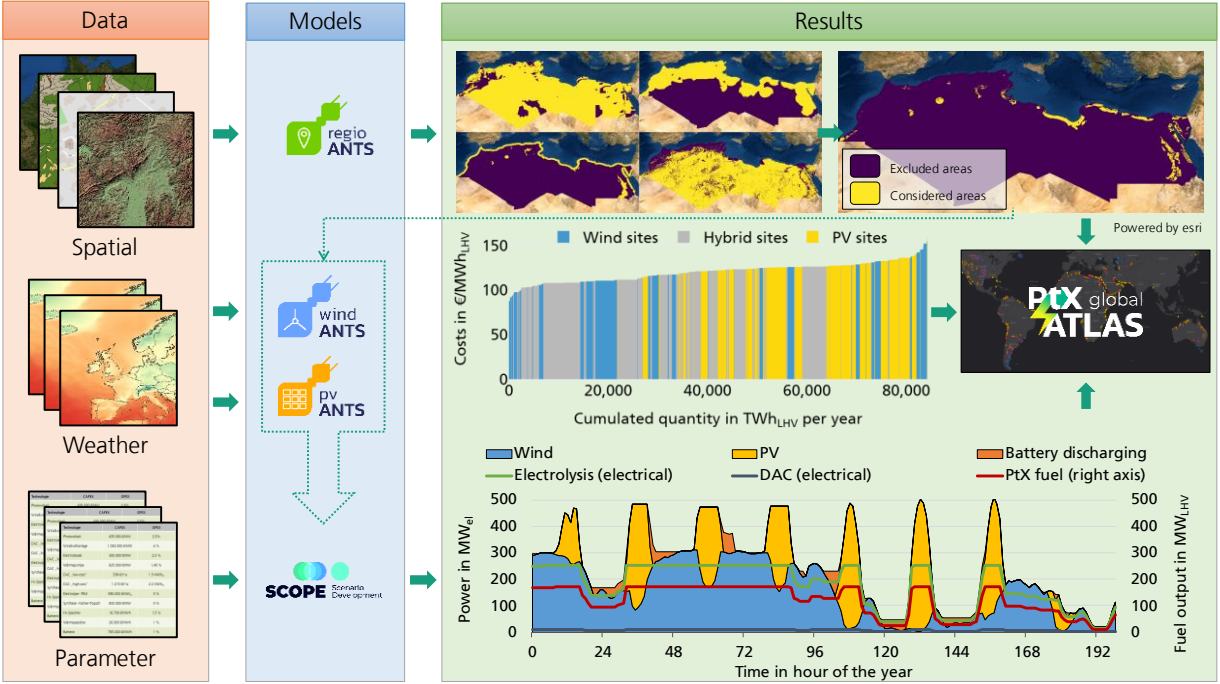

# 1. Role of PtX fuels in the energy transition

In response to the Paris Agreement of 2015 [1], the European Commission's ambition is to make Europe the first climate-neutral continent by 2050 [2]. This requires a comprehensive conversion of the energy supply system to replace the use of fossil fuels. Today, fossil fuels are imported in large quantities. In the future, these energy quantities should no longer come from fossil fuels, but a supply without imports from non-European regions nevertheless seems unavoidable, as [3] already expects for the year 2030.

Electricity-based fuels[2] from renewable energies are to replace fossil fuels where the direct use of electricity from renewable energies is not possible. Steel manufacturers, for example, can use hydrogen instead of coal. Synthetic fuels may replace kerosene or heavy oil and diesel in aviation and shipping. Dispatchable power plants (e.g. gas turbines) needed for a secure power supply can be operated with synthetic natural gas or hydrogen instead of fossil fuels. The synthetic production of electricity-based fuels enables a net-neutral decarbonisation of these sectors and thus offers a major opportunity to implement an almost greenhouse gas-neutral energy system.

Since the production of Power-to-X (PtX) fuels is very energy-intensive and involves high losses even with future efficiency improvements, production only appears to make economic sense in regions with favourable renewable energy resources and corresponding land availability. For countries that have such potential in large quantities and beyond their own needs, PtX fuels open the door to an attractive future market in the transformation process. For the transportation sector alone, global PtX requirements are assumed to be at least 10,000 TWh/yr [4] up to 35,000 TWh/yr [5] and even higher, as fossil fuel industry stakeholders such as [6] claim. For Europe, [7] assume between 792 TWh/yr and 1,782 TWh/yr for the transport sector and up to 3,102 TWh/yr if all sectors are considered. For Germany, an evaluation of the "Big 5" climate neutrality studies in [8] shows a demand for PtX between 215 TWh/yr and 657 TWh/yr.

In recent years, the analysis of potentials and future costs of PtX fuels has received increasing attention, and several studies have been published. These studies vary considerably in their regional and temporal scope, the technologies and PtX products considered. Table 1 gives an overview of the most important characteristics found in the literature.

The temporal scope of the related literature spans from the short- to the long-term perspective. Some studies even model pathways from today to 2050. The regional scope is also quite heterogeneous. Some studies focus on single countries, see [9] or [10] for Argentina, while others conduct the analysis on a global level [11] or [12]. Many authors put the focus on promising regions such as the MENA[3] region (e.g. [13], [14], [15] or [16]).

Concerning the power sources for the PtX fuels, all studies consider at least wind power or photovoltaics (PV), sometimes also combined as hybrid plants. Only a few authors [11,17–20] include battery storage systems as an additional flexibility measure for the fluctuating electricity output. Concentrated solar power (CSP) is only considered in two cases [13,16] while geothermal energy and hydropower are analysed in three studies [16,21,22].

The electrolyser technology considered by previous studies also varies. While a few studies analyse three electrolyser technologies, i.e. alkaline electrolysis - AEL, proton exchange membrane – PEM, and solid oxide electrolysers - SOEC [15,16,23,24], most authors focus only on one technology.

A considerable share of the studies analyse only the potential of hydrogen [9–12,17,18,20,23–27] while others have a broader scope and also consider methane, methanol, ammonia, and/or Fischer-Tropsch fuels.

---

[2] Electricity-based fuels ("E-fuels") or PtX fuels is an umbrella term for all gaseous and liquid energy carriers produced from renewable electricity considered in this analysis.
[3] Middle East and North Africa

Table 1: Overview of related literature analysing PtX potentials and costs (Source: Own illustration)

| Source | Time scope | Regional scope | Power sources | Electrolysis technology | PtX products |
|---|---|---|---|---|---|
| [9] | "Present case", "future case" | Argentina | wind | not mentioned | hydrogen |
| [10] | not mentioned | Argentina | wind | PEM | hydrogen |
| [11] | 2020, 2030, 2040, 2050 | Whole world | PV, wind, battery | not mentioned | hydrogen |
| [12] | "long term" | Whole world | PV, wind (onshore) | not specified | hydrogen |
| [13] | 2030, 2050 | MENA region (Morocco, Algeria, Tunisia, Libya, Egypt, Israel, Lebanon, Turkey, Syria, Jordan, Saudi Arabia) | PV, wind (on- and offshore), CSP | PEM, SOEC | hydrogen, synthetic methane |
| [14] | 2030, 2040 | Maghreb region (Morocco, Western Sahara, Algeria, Mauritania, Tunisia, Libya) | PV, wind | AEL | synthetic methane, Fischer-Tropsch diesel |
| [15] | 2020, 2030, 2040, 2050 | MENA region | PV, wind | AEL, PEM, SOEC | H2, synthetic methane, methanol, Fischer-Tropsch fuel |
| [16] | 2015, 2050 | Europe, North Africa and the Middle East (EUMENA) | PV, wind, geothermal energy, hydropower, CSP | AEL, PEM, SOEC | hydrogen, synthetic methane, methanol, Fischer-Tropsch fuel |
| [17] | 2020 | North Africa | PV, wind, battery | PEM | hydrogen |
| [18] | 2030, 2050 | 30 Non-EU countries around the world | PV, wind, battery | not mentioned | hydrogen |
| [19][4] | 2050 | Morocco, Tunisia | PV, wind, battery | PEM | hydrogen |
| [20] | 2020, 2030, 2040, 2050 | Europe, Northern Africa, Middle East | PV, wind (onshore), battery | PEM | hydrogen |
| [21] | 2035 | Chile, Argentina, Canada, Iceland, Namibia, Egypt, Australia | PV, wind, geothermal energy, hydropower | AEL, PEM | hydrogen, methanol, Fischer-Tropsch diesel |
| [22] | 2020, 2030, 2050 | North and Baltic Seas, Iceland, | PV, wind (on- and offshore), geothermal | low-temperature | synthetic methane, methanol, Fischer-Tropsch fuel |

---

[4] In this study a former analysis of the authors was published. In contrast to the present study only individual sites and PtX supply pathways were considered.

| Source | Time scope | Regional scope | Power sources | Electrolysis technology | PtX products |
| --- | --- | --- | --- | --- | --- |
| | | North Africa, Middle East | energy, hydropower | electrolysis (not specified) | |
| [23] | 2018, 2050 | Northern Africa | PV, wind | AEL, PEM, SOEC | hydrogen |
| [24] | 2020, 2030, 2050 | EU and non-EU-countries | PV, wind (on- and offshore) | AEL, PEM, SOEC | hydrogen |
| [25] | 2050 | Strong wind and solar regions around the world | PV, wind | not specified | hydrogen |
| [26] | 2020 to 2050 | 94 countries on six continents (except Antarctica) | PV, wind (on- and offshore) | low and high temperature electrolysis | hydrogen |
| [27] | 2030, 2050 | Whole World | PV, wind (on- and offshore) | AEL | hydrogen |
| [28] | 2030, 2040, 2050 | Germany, Denmark, Spain, Morocco, Egypt, Saudi Arabia, Argentina, Australia | PV, wind (on- and offshore) | AEL | hydrogen, methane, methanol, ammonia, Fischer-Tropsch fuels |
| [29] | 2020, 2030 | Morocco/ Western Sahara | PV, wind | PEM | hydrogen, methane, methanol, ammonia |
| [30] | 2020-2025, 2030, 2050 | Australia, Northwest Africa | PV, wind | "median of AEL/PEM" | hydrogen, synthetic methane, synthetic fuels (diesel, kerosene) |
| [31] | 2050 | 10 windy and 15 sunny regions around the world | PV, wind | PEM | hydrogen, synthetic methane, synthetic fuels |
| [32] | 2020, 2030, 2040, 2050 | global analysis (numbers given on continent level) | "power from the system" | AEL | hydrogen, synthetic methane, Fischer-Tropsch fuel |

This paper contributes to the existing literature by a comprehensive analysis of the production of electricity-based fuels from onshore wind energy and ground-mounted photovoltaic energy for all countries and regions outside the European Economic Area along 14 PtX production pathways. These include hydrogen (gaseous and liquid), methane (gaseous and liquid), methanol, Fischer-Tropsch fuels and ammonia each with two different electrolyser technologies (PEM and SOEC). An optimisation model designs cost-optimal system configurations of electrolysis, synthesis, heat source, wind and solar plants, and storage systems (battery, heat, hydrogen, methane) for each site based on the local conditions. The technical and economic potential assessment uses data with high temporal (1 hour) and spatial (1 km) resolution. A particular focus is on the consideration of available land and prevailing weather conditions, and factors such as local water availability, nature conservation aspects or distance to infrastructure.

PtX products are suitable for decarbonisation of the domestic energy system and furthermore also for export. As European countries will have a future import demand for these products, we investigate socio-economic factors in the producing countries and import options to Europe for further evaluation. The socio-economic analysis is based on indicators and associated indices from literature. For the transport cost calculation we use a detailed transport model and illustrate the results in this paper for the case of Germany as an importing country.

## Outline

The remainder of this paper is organised as follows: Section 2 explains the methodology, including the modelling and optimisation approach to derive cost-optimal production systems for PtX fuels. Section 3 sets out the data and assumptions of the study. Section 4 presents and section 5 discusses the results. The paper closes with a summary and relevant conclusions in section 6.

## 2. Methodology

The objective of this analysis is to estimate the long-term production costs and generation quantities of PtX products for all countries and regions outside the European Economic Area. We focus on seven different synthetic fuels as final products:

- Gaseous hydrogen
- Liquid hydrogen
- Liquid methane (liquid natural gas, LNG)
- Gaseous methane (compressed natural gas, CNG)
- Methanol
- Fischer-Tropsch (FT) fuels
- Ammonia

As electricity source, we consider either onshore wind power, ground mounted photovoltaic energy or a combination of both.

Several data sets and models were used for the study. Fig. 1 shows a schematic overview of the methodology.

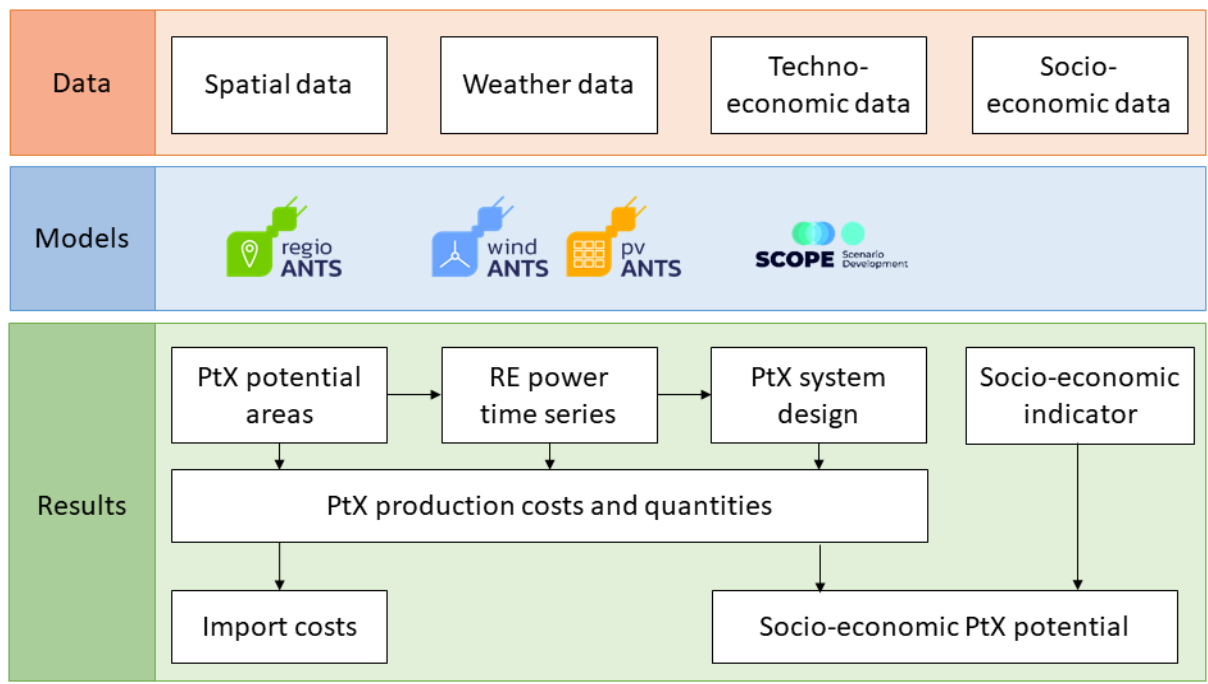

*Fig. 1: Overview of applied methodology. Input data indicated in orange, applied models in blue and results in green (RE stands for renewable energies; Source: Own illustration).*

The data basis includes high-resolution spatial and weather data as well as techno-economic and socio-economic parameters. The energyANTS model family is used to identify potential areas (model "regioANTS") and simulate power generation time series for onshore wind plants (model "windANTS") and ground-mounted photovoltaic plants (model "pvANTS"). The energy system optimisation model SCOPE SD is used to determine a cost optimal system design for 14 different PtX production pathways (seven final products times two electrolysis technologies). In addition to that we present a detailed transport cost calculation and a socioeconomic analysis to show import options for the analysed PtX products to Europe. A detailed description of the methodological approach is given in the following subsections.

## 2.1. Area Identification and selection of suitable locations

The potential area analysis is based on a Boolean superposition of various exclusion criteria, i.e., factors that rule out land use for PtX generation. These criteria are defined within a criteria catalogue and applied using globally available spatial data. The potential area analysis uses a grid resolution of 1 km and is carried out individually for each country. Fig. 2 gives an overview of the approach.

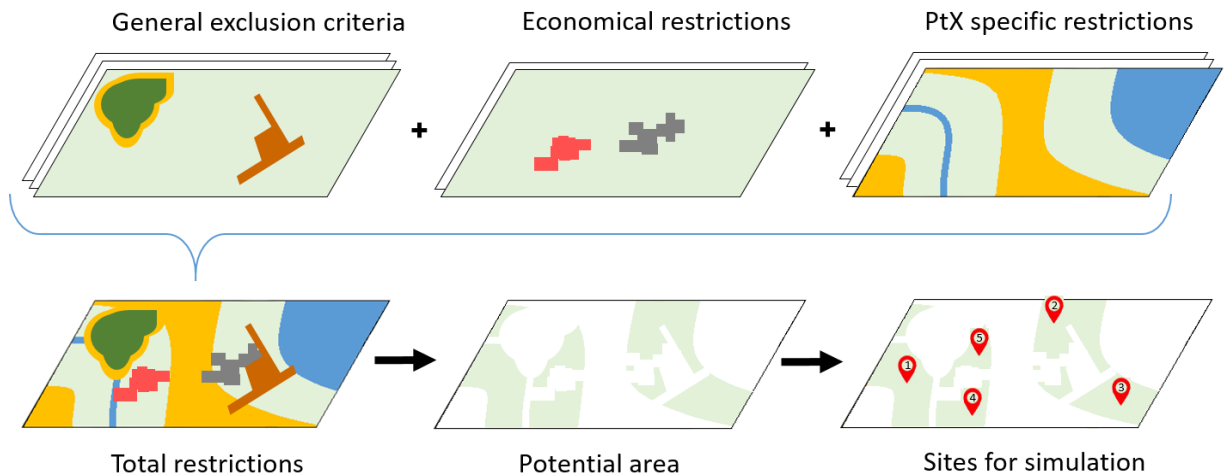

Fig. 2: Illustration of the identification and selection of suitable locations for PtX production (Source: Own illustration).

### 2.1.1. PtX potential area identification

The analysis of PtX potential areas can be broken down into three steps. In the first step, generally applicable criteria for excluding land for wind or PV power plant suitability are defined. In the second step, economic restrictions are applied to the areas. In the third step, the analysis assigns technical and ecological restrictions with regard specifically to PtX technologies for the remaining areas. Here, two variants are distinguished, differing for the assumed water source (coastal or inland water).

A rough estimate of future land potentials is based on universal criteria for potential area analysis for renewable energies (RE). General exclusion criteria are defined for land use, such as built-up areas, population densities or agricultural land or forest (cf. Table 2). The analysis also considers nature conservation areas (both onshore and offshore) with a buffer of 1 km and slope gradients in order to exclude areas that are too steep for building up RE capacity. Areas with a mean slope (in a 1 km grid) greater than 5° are excluded.

The economic restriction that is taken into account is the levelized cost of electricity (LCOE) for both RE sources. The LCOE are calculated based on techno-economic assumptions for reference plants[5]. LCOE of 30 €/MWh for photovoltaics and 40 €/MWh for wind turbines are set as the upper limit of LCOE for suitable areas. In addition to pure photovoltaic and pure wind sites where the LCOE is lower than the cost limit of the corresponding technology, hybrid sites constitute another category where the LCOE of both generation technologies is lower than the cost limit of each technology.

The criteria used explicitly for examining the suitability of PtX fuel production are mainly distance criteria. Firstly, the availability of skilled labour, such as engineers, must be ensured for large-scale PtX projects. We assume that this criterion is fulfilled in "larger" cities (hereafter referred to as "city") and therefore a distance of less than 200 km to cities is set as a further criterion. Furthermore, these cities can serve as a purchaser or consumer for the produced PtX fuels and thus additionally enhances the suitability of the nearby generation sites. Secondly, the distance to the nearest distribution infrastructure, in particular ports and pipelines, is considered to ensure the possibility to distribute the PtX fuels nationally and internationally.

---
[5] cf. Table 7 and Appendix B. 3

The selection criterion is set to be either closer than 50 km to a pipeline (for gaseous PtX products like hydrogen or methane) or 500 km to a port (for liquid PtX products including Fischer-Tropsch fuel, methanol or ammonia) to be included. As water electrolysis requires a constant supply of feed water, the third infrastructural criterion is the distance to the nearest water access point, i.e. seawater for coastal sites and freshwater for inland sites. All areas further than 50 km to these access points are excluded.

The technological criteria are combined with two ecological ones. Inland sites must have a low water stress level. Whereas for coastal sites we exclude land along marine protection areas to prevent adverse environmental impacts caused by the sole from desalination plants [33].

Subsection 3.1 provides more information on the datasets that were used during the potential area analysis.

### 2.1.2. Site selection for detailed PtX analysis

From the GIS-based potential area analysis, the most suitable sites are selected as preparation for the downstream analysis steps for the RE yield estimation. The selection of these most suitable sites is based on the size of spatially contiguous grid areas that were identified as potential areas. The assumption is that large-scaled PtX production facilities will first be built on as large an area as possible. Only spatially contiguous areas larger than 10 km² are considered for the site selection. All previously defined categories (pure wind, pure PV, and hybrid sites on coastal or inland waters) are considered, with up to five sites selected. In total, up to 30 sites are thus selected for each country. This results in almost 600 representative sites around the globe where the following steps for the yield estimation for RE and the expansion and deployment optimisation for 14 PtX production pathways are carried out.

## 2.2. Yield estimation for renewable energies

The determined suitable areas for PtX generation serve as a basis for the simulation of temporally high-resolution generation time series for electricity from onshore wind and ground-mounted PV systems. The electricity yields simulated with the models windANTS and pvANTS uses weather information from the ECMWF's ERA5 weather model [34]. The temporal resolution of the RE yield simulations is one hour and we performed it for five historical weather years (2008 to 2012).

According to this methodology, five time series for both of the RE sources were simulated for every ERA5-model-pixel that overlays with any of the nearly 600 selected sites. An aggregation takes place if more than one ERA-model-pixel is overlaying with the corresponding area of this site. Depending on the area of this overlay, the time series of the individual pixels are combined according to Equation (2-1):

$$P_{norm,ges} = \sum_{i=0}^{n} P_{norm,i} \cdot \left(\frac{A_i}{A_{ges}}\right) \qquad (2\text{-}1)$$

Here, $P_{norm,ges}$ is the aggregated capacity factor time series of a site, $n$ denotes the number of ERA5-model-pixels with which this site overlays, $P_{norm,i}$ is the normalised time series in the respective ERA5-model-pixel, $A_i$ is the area of the overlay and $A_{ges}$ is the total area of the analysed site.

For further calculation, one year as a representative weather year is selected out of the aggregated time series based on the average yield (over the five analysed years) per site following [35].

### 2.2.1. Wind energy

The simulations of the generation time series for onshore wind are carried out with the submodel windANTS from the model family energyANTS. A detailed description of the physical wind model can be found in [36]. Fig. 3 gives an overview of the used parts of the model.

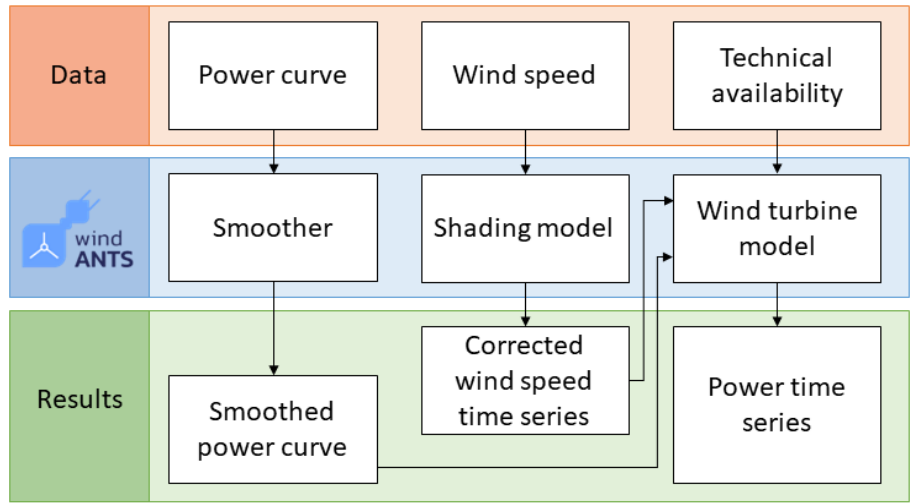

Fig. 3: Overview of physical wind model. Input data shown in orange, used parts of the model in blue and results in green (Source: Own illustration).

The characteristics of two different turbine types are used. These differ in hub height (180 m and 130 m) and specific area power (250 W/m² and 300 W/m²). For each type, a synthetic power curve is generated and smoothed. The smoothing of the power curve is necessary because the wind speeds of the weather model represent a temporal and spatial average. Furthermore, the wind speeds are interpolated to the hub height and corrected due to shading effects within a wind park. Finally, the power time series are simulated for each location considering the smoothed power curve, the corrected wind speed and the technical availability of the wind turbines.

### 2.2.2. Photovoltaic energy

The simulations of the generation time series for ground mounted photovoltaic energy are carried out with the submodel pvANTS. A detailed description of the physical PV model can be found in [36]. Fig. 4 gives an overview of the used parts of the model.

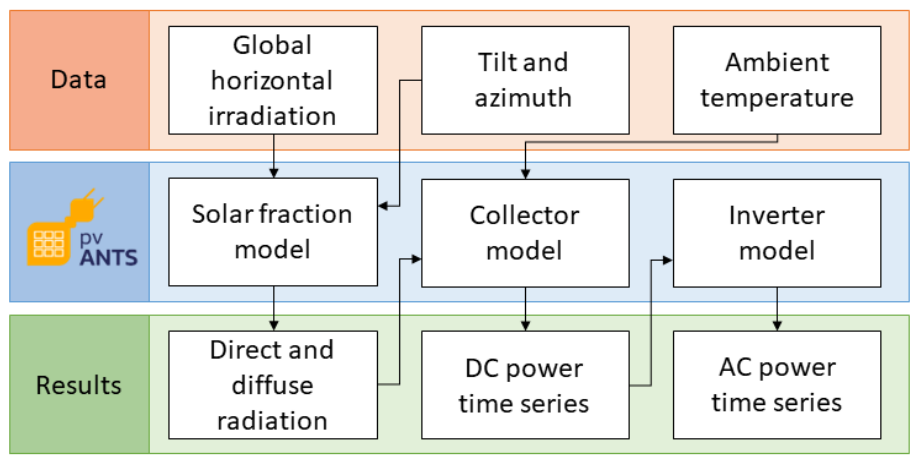

Fig. 4: Overview of physical PV model. Input data shown in orange, used parts of the model in blue and results in green (DC - direct current, AC - alternating current; Source: Own illustration).

For the power time series of PV, the optimal tilt angle and the best orientation from [37] are fed to the solar fraction model which uses the module orientation, the global horizontal irradiation as well as the sun position from [38] to compute the direct and diffuse radiation on the module plane. The collector model then generates a direct current (DC) power time series with a temperature dependent efficiency function. In the last step, the DC time series is converted to alternating current (AC) using the inverter model. Several losses (dirt on modules, cabeling losses etc.) are included as well.

## 2.3. Modelling of site specific PtX system configurations

In order to derive the generation costs for different synthetic fuels on different locations, the model SCOPE SD of Fraunhofer IEE is used [39]. With the help of this model, the cost-optimal composition of a system of generation technologies, processing and storage units are evaluated under the local weather conditions.

Depending on the final product and electrolyser type (PEM or SOEC), the system may consist of different components. In each case, power generation comes from wind power and/or solar PV where time series for the potential site-specific power generation are taken from the results of the windANTS and pvANTS models (see subsection 2.2). Furthermore, a battery storage can be used to provide additional flexibility. In case of high temperature electrolysers, heat comes from an electric boiler, potentially combined with a thermal storage system. Produced hydrogen can be stored in a hydrogen storage before being further processed, i.e. synthesis, liquefaction, or compression. The storage can also have a capacity of zero, implying that it is only a balancing point. Carbon dioxide ($CO_2$) from direct air capture (DAC) or nitrogen from an air separating unit (ASU) is required for the synthesis process. The waste heat of the synthesis can be partly used by the electrolyser unit (SOEC) or, in the case of the PEM production pathways (except the Haber-Bosch-process), for the DAC unit. Produced methane can also be stored in a methane storage, if desirable. Fig. 5 gives an overview of the components and processes in the PtX model with their energy and $CO_2$ flows.

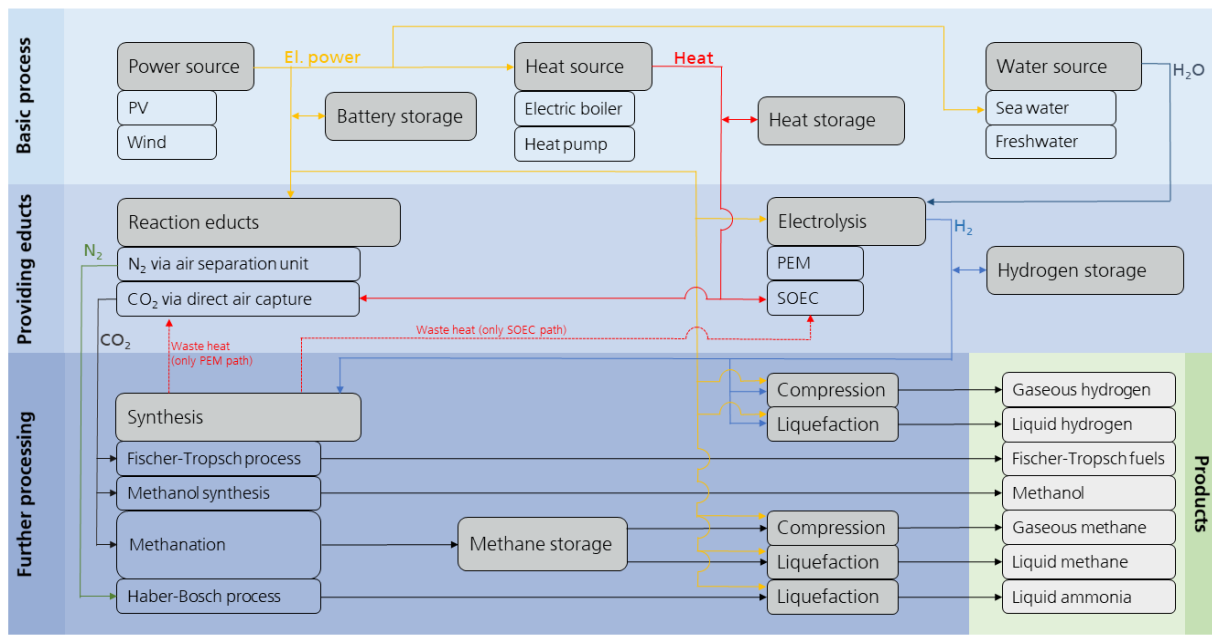

*Fig. 5: Overview of the PtX system model (Source: Own illustration).*

The linear cost optimisation model SCOPE SD considers both the investment and the variable and fixed operating costs for each system component. A detailed mathematical model description can be found in "A Appendix SCOPE SD model".

## 2.4. Total PtX generation quantity

Based on the area identification and system design optimisation, we derive an estimation of future PtX fuel generation quantities and fuel costs per quantity for all countries in the case study. The determination of the PtX area requirements is based on the area requirements of the RE generators, taking into account additional requirements for further PtX system components such as electrolysis, synthesis, liquefaction, DAC or ASU technologies and storage. Accordingly, an aggregated area requirement of 15 $MW_{wind}$/km² is assumed for PtX from wind energy and 40 $MW_{PV}$/km² for PtX from PV energy. For hybrid sites, a combined land use for both wind and PV power generation is assumed, i.e. wind energy is the limiting factor and 15 MW/km² is the capacity-specific land use density. The resulting PtX output energy depends on the cost-optimised system design and varies across locations.

In the case of an overlay of potential sites that have access to both, inland waters and coastal waters, the use of an Analytical Hierarchical Process (AHP) assessment method allows the sites to be assigned to only one category. During the AHP, we select seven criteria that mainly determine the quality of a site in terms of PtX fuel production. Subsequently, we weight these criteria and we introduce interval limits, each of which can be assigned to a quality class between 0 (excluded) and 5 (best). With a combination of these two methods, we calculate an overall AHP-score between 0 and 5 for every pixel, representing the pixel's quality in terms of PtX fuel production. The chosen criteria and their evaluation weights are different for every category, which is why one pixel can have different AHP-scores (in the case of an overlay). Every pixel is assigned to the category in which it has the highest AHP-score. This is necessary to consider each potential area only once when calculating the cumulative generation potential. If the AHP ranking does not generate a clear assignment to a category, the distance to the nearest water access (coastal or inland) is determined as the final decision criterion. The potential is assigned to the category with the shorter distance.

For potential areas resulting from the PtX area identification that are not selected for detailed PtX analysis (cf. subsection 2.1.2), a simplified scaling approach is chosen for the determination of quantities and costs. The energy generation quantities and production costs of the simulated sites per site type, meaning per RE technology (wind/PV/hybrid) per category (coastal, inland) and per country, are transferred to the remaining areas. For this purpose, a scaling factor per type is calculated and applied based on the ratio between the potential area per site type and the area of the simulated sites.

## 2.5. Transport cost model

The determination of the transport costs of synthetic fuels is based on a cost model for tanker ships, which calculates the transport costs in the unit €/MWh$_{LHV}$ for the liquid fuel variants (Fischer-Tropsch fuel, methanol, liquid methane, liquid hydrogen and ammonia) depending on the distance between the importing and exporting country. To calculate the distance, the largest port of the country is taken into account in each case. If the country does not have a port, the nearest foreign port is selected. We use the MARNET data set from [40] to calculate the distance to be covered. As a result, distance-dependent transport costs per unit of energy are calculated, which, in conjunction with the production costs, represent the energy supply costs in the respective importing country.

For the calculation of the transport costs, we assumed the transport via a reference ship with a total deadweight tonnage (DWT) of 280,000 tons. Depending on the gravimetric and volumetric energy density of the fuel to be transported and taking into account the transport tanks, we calculate a fuel-specific possible transport capacity. In order to obtain framework conditions for the overseas transport of energy carriers in the course of the calculation model, we utilise influencing factors based on [41]. In general, it is assumed that the transport ships are powered by the respective fuel to be transported. A hydrogen tanker, for example, will have a propulsion system consisting of a polymer membrane fuel cell and an electric motor including the associated power electronics. The technology-specific costs for the corresponding propulsion systems are included in the investment costs, which are incorporated in the annual fixed costs of the tanker service using the net present value method. For this purpose, a depreciation period of 27 years is applied. Furthermore, the fixed operating costs, mainly consisting of maintenance costs, costs for insurance and administration as well as provision of reserve capacity, which is needed to compensate for the maintenance times of the tanker, are considered. The variable operating costs, in particular the fuel consumption and the boil-off (evaporation) of the liquefied gas variants, are also considered. In these variants we assume the direct use of the evaporating liquefied gases as ship fuel.

The most important technologic and economic assumptions for the calculation of the normalised transport costs according to [41] in €/(MWh$_{LHV}$ × km) are shown in Table 8 in Subsection 3.3.

We multiply these distance dependent values with the distance between the largest port in each production country and the importing port in the corresponding import country to derive the transport costs per MWh$_{LHV}$ for each fuel. This consideration allows a better comparison of the different end products along possible import routes.

## 2.6. High-level socioeconomic analysis

The suitability for the development of a PtX infrastructure is also dependent on the socioeconomic conditions in the PtX generating countries. Using the method of a global high-level analysis, the socioeconomic potentials of countries and regions are considered based on the following thematic fields:

Economy, politics, society, technology, and natural conditions. In order to evaluate the question from Germany's point of view, the topic area of proximity to Germany is also included. It is considered that not only the geographical proximity is decisive for a possible investment, but also the existence of a developed logistics infrastructure in the exporting country or existing economic relations between a possible PtX exporting country and Germany.

In the high-level analysis, the average value from the six topic areas yields the socioeconomic potential of a PtX exporting country. Each of these thematic fields is based on indicators and associated indices. For example, the topic area "society" is made up of the indicators unemployment, satisfaction and peace, health care system, population, climate change effects and energy demand. In the analysis, the individual values of a total of 40 indicators and more than 70 associated indices were used in the six thematic areas. The analysis was conducted in 2020-2021 and is based on datasets from international organizations (including the World Bank; OECD) and private firms. More recent crisis in the world (e.g. the Russo-Ukrainian war) were not taken into account in the underlying data sets. The exact methodological procedure can be found in [42].

## 3. Data

In this chapter, the most important data used for the analysis is presented.

### 3.1. Data for potential area analysis

For each criterion that was defined in subsection 2.1, a suitable GIS dataset is selected. The presentation of the datasets that are used follows the methodology and is divided into three parts comprising general, economical and PtX specific data.

#### 3.1.1. Data on general area identification

The dataset that represents the administrative boundaries for each country is the "level 0" of the Database of Global Administrative Areas (GADM) dataset, which depicts the outer national boundaries as a polygon [43]. Since the exclusion of areas worthy of protection must be guaranteed, we regard two datasets for this purpose. One is the World Database on Protected Areas (WDPA, version 07/2019), which contains all land and water areas worthy of protection according to specific criteria [44]. The second one is the Global Critical Habitat Screening Layer, which includes all likely or potentially critical habitats on land and in the sea [45]. To identify the land uses that shall be excluded from the consideration (cf. Table 2) the European Space Agency (ESA) Land Cover is applied, which provides information on land use in a global context [46]. We utilise a dataset of the world's populated places from maxmind [47] with a buffer of 1 km to refine the data on settlement areas. Furthermore, we consider the population density with the Gridded Population of the World (GPW) v4 dataset, which contains information on the population density in a global context [48]. To ensure the exclusion of croplands, we also use the dataset Global Food Security-Support Analysis Data [49]. As a basis for the calculation of the slope (cf. subsection 2.1) we apply the data from the Shuttle Radar Topography Mission (resolution of 3 arc seconds), which form a high-resolution digital terrain model [50].

#### 3.1.2. Economic data for photovoltaics and wind turbines

The economical assessment based on the LCOE calculation of the RE sources. For this, we use data from the Global Wind Atlas and the Global Solar Atlas [37,51], which both provide long-term average values to estimate average production quantities. We use weibull factors to assess the wind potential and the prefabricated Photovoltaic power potential for just that.

#### 3.1.3. PtX specific datasets

We use the World Cities Database in its basic version [52] as a basis for calculating the distance to the nearest city. It contains about 41,000 cities that are assumed to fulfil the conditions that we declared in subsection 2.1. To calculate the distance to distribution points (ports and pipelines) we use three datasets. The World Port Index [53] includes most of the world's ports as point data for the port's distance. Note that the smallest port size ("very small") is not taken into account for the calculations since this harbour size does not fulfil the requirement of being capable to distribute the produced PtX fuels. For the distance to pipelines, two data sets on pipelines in North America [54] as well as the Eurasian region and North Africa [55] are used.

For coastal sites, the distance to the coast is calculated by using the GADM level 0 dataset, a dataset that contains the Exclusive Economic Zones of each country with seawater access [56] and the marine protected areas from the WDPA dataset. Whereas for the inland sites we utilise the ESA Global Water Bodies dataset [57] as reference for the nearest freshwater access point. The Aqueduct Water Risk Indicators 3.0 [58] help to take the criterion of water stress into account.

A summary of all criteria employed during the analysis and their corresponding source is given in Table 2.

*Table 2: Catalogue of criteria for the identification of suitable locations for the production of PtX*

| Criteria | Exclusion criterion | Argument | Source |
|---|---|---|---|
| General | Land use | Forests, built up area, cropland, water bodies, snow and ice areas | [46,49] |
| General | Slope | >5° (1 km resolution) | [50] |
| General | Settlement areas | All settlement areas with a buffer of 1 km | [46,47] |
| General | Population density | > 50 inhabitants/km² | [48] |
| General | Protected areas | Nature and landscape conservation as well as potentially critical habitats with a buffer of 1 km | [44,45] |
| Economic | LCOE wind | > 40 €/MWh | [51] |
| Economic | LCOE photovoltaics | > 30 €/MWh | [37] |
| PtX specific | Distance to ports | > 500 km | [53] |
| PtX specific | Distance to pipelines | > 50 km | [54,55] |
| PtX specific | Distance to cities | > 200 km | [52] |
| PtX specific | Distance to the national coastline | > 50 km | [43,56] |
| PtX specific | Marine protected areas | Coastline along marine protected areas with a buffer of 4 km | [44] |
| PtX specific | Distance to inland water source | > 50 km | [57] |
| PtX specific | Water stress | > low | [58] |

## 3.2. Weather data for modelling energy production time series with high temporal resolution

As an input for the simulation of generation time series for both RE sources, we extract weather data from the ERA5 weather model of the European Centre for Medium-Range Weather Forecasts [34]. It provides extensive weather data from 1950 to present with a temporal resolution of one hour and a spatial resolution (pixel size) of approx. 31 km × 31 km. The extracted meteorological data includes, among other parameters, the solar radiation (downwards) at a surface level, the temperature 2 m above surface level as well as the u- and v- components of the wind (speed) at the corresponding hub heights (cf. subsection 2.2).

## 3.3. Techno-economic parameters for cost analysis

The estimation of the PtX fuel production costs is based on an investment and dispatch optimisation with the optimisation model SCOPE of Fraunhofer IEE. In the following, we describe simulation assumptions based on hourly resolved weather data for the scenario year 2050. All calculations are carried out with a cost of capital assumed at 8 % [59].

Because the calculations take place for the year 2050, there is a high degree of uncertainty in the techno-economic assumptions. In the literature, there are rather optimistic and rather pessimistic ones (cf. section B. 4 of the Appendix). We attempt to use average values from these sources. For all system components for which the variable operating costs ($OPEX_{var}$) are not explicitly listed, these are included in the fixed operating costs ($OPEX_{fix}$). Table 3 contains the techno-economic assumptions for both electrolyser options. The value of the operating costs is stated as a percentage of the investment costs (CAPEX) per year.

Table 3: Techno-economic assumptions of the PEM electrolyser [own assumptions after60] and the SOEC electrolyser [60,61]

| Technology | CAPEX | $OPEX_{fix}$ | Efficiency | Heat demand |
|---|---|---|---|---|
| PEM electrolyser | 470,000 €/$MW_{el}$ (input) | 5 % CAPEX | 71.0 % (overall) | - |
| SOEC electrolyser | 550,000 €/$MW_{el}$ (input) | 5 % CAPEX | 88.0 % (electrical) | 44.3 kJ/$mol_{H2}$ (output) |

The techno-economic assumptions of the three different synthesis process types are shown in Table 4. Due to the uncertainties of the cost development until 2050, the investment costs of the synthesis technologies were not varied following [62]. The efficiency was calculated using the chemical reaction of the synthesis process with procedural losses of 5 % (e. g. electrical consumption of auxiliary devices). Concerning the recovery and use of waste heat, a heat exchanger efficiency of 50 % is assumed.

Table 4: Techno-economic assumptions of the FT synthesis, the methanol synthesis, the methane synthesis and the ammonia synthesis [61,own calculations, 63,own assumptions after64–66]

| Technology | CAPEX | $OPEX_{fix}$ | Efficiency | Waste heat (usable) |
|---|---|---|---|---|
| FT synthesis | 324,000 €/$MW_{H2}$ (input) | 5 % CAPEX | 76.3 % (overall) | 57.9 kJ/$mol_{CH2-chain}$ (output) |
| Methanol synthesis | 324,000 €/$MW_{H2}$ (input) | 5 % CAPEX | 79.1 % (overall) | 24.9 kJ/$mol_{CH3OH}$ (output) |
| Methane synthesis | 324,000 €/$MW_{H2}$ (input) | 5 % CAPEX | 78.9 % (overall) | 82.5 kJ/$mol_{CH4}$ (output) |
| Ammonia synthesis | 507,000 €/$MW_{H2}$ (input) | 2 % CAPEX | 83.02 % (overall) | 23.1 kJ/$mol_{NH3}$ (output) |

The techno-economic assumptions of the compression and liquefaction for methane and hydrogen can be seen in Table 5.

Table 5: Techno-economic assumptions of the methane and hydrogen compression and liquefaction [67,68,own assumptions after69]

| Technology | CAPEX | $OPEX_{fix}$ | Power consumption |
|---|---|---|---|
| Methane compression | 3,900 €/$kW_{el}$ (input) | 4 % CAPEX | 0.03 $kWh_{el}$/$kWh_{CH4}$ |
| Methane liquefaction | 500 €/($t_{CH4}$·yr) (output) | 4 % CAPEX | 0.08 $kWh_{el}$/$kWh_{CH4}$ |
| Hydrogen compression | 3,900 €/$kW_{el}$ (input) | 4 % CAPEX | 0.048 $kWh_{el}$/$kWh_{H2}$ |
| Hydrogen liquefaction | 3,500 €/($t_{H2}$·yr) (output) | 4 % CAPEX | 0.2 $kWh_{el}$/$kWh_{H2}$ |

The techno-economic assumptions of the DAC and the ASU-unit are shown in Table 6.

Table 6: Techno-economic assumptions of the DAC and the ASU-unit [own assumptions after 15,29,32,66,70–72]

| Technology | CAPEX | OPEX$_{fix}$ | OPEX$_{var}$ | Power consumption | Heat demand |
|---|---|---|---|---|---|
| DAC unit | 450 €/(t$_{CO2}$·yr) (output) | 4 % CAPEX | 1.30 €/MWh$_{el}$ | 255.15 kWh$_{el}$/t$_{CO2}$ | 1,312.2 kWh$_{th}$/t$_{CO2}$ |
| ASU unit | 165 €/(t$_{N2}$·yr) (output) | 2 % CAPEX | - | 100 kWh$_{el}$/t$_{N2}$ | - |

Table 7 contains the techno-economic assumptions of the remaining most important technical components of the PtX supply system. Additionally, the variable operating costs for the electric boiler are assumed to be 0.40 €/MWh$_{el}$ and 1.69 €/MWh$_{el}$ for the large-scale heat pump.

Table 7: Techno-economic assumptions of the renewable energy sources, the heat technology aggregates and the storage technologies [35,own assumptions after 73–75]

| Technology | CAPEX | OPEX$_{fix}$ |
|---|---|---|
| Wind power plant with 180 m hub height | 886,000 €/MW$_{el}$ (output) | 4 % CAPEX |
| Wind power plant with 130 m hub height | 806,000 €/MW$_{el}$ (output) | 4 % CAPEX |
| Photovoltaic plant | 321,000 €/MW$_{el}$ (output) | 2.5 % CAPEX |
| Large heat pump | 1,011,000 €/MW$_{el}$ (input) | 1.45 % CAPEX |
| Electric boiler | 100,000 €/MW$_{el}$ (input) | 2.5 % CAPEX |
| Methane storage | 5,015 €/MWh$_{CH4}$ (capacity) | 1 % CAPEX |
| Hydrogen storage | 16,700 €/MWh$_{H2}$ (capaciyt) | 1.5 % CAPEX |
| Heat storage | 26,000 €/MWh$_{th}$ (capacity) | 1 % CAPEX |
| Battery storage | 479,500 €/MWh$_{el}$ (capacity) | 1 % CAPEX |

Transport cost

The data that we use for calculating the distance dependent transport cost can be seen in Table 8. For travelling speed we assume 16 knots (kn), also called "super slow steaming", because this significantly decreases fuel consumption.

Table 8: Techno-economic assumptions for the calculations of the fuel transport based on [41]

|  | FT fuel | Methane (liquid) | Hydrogen (liquid) | Ammonia | Methanol |
|---|---|---|---|---|---|
| Efficiency driving unit (%) [76–78] | 55 | 55 | 47 | 49 | 55 |
| Carrying capacity (DWT) [41] | 280,000 | 280,000 | 280,000 | 280,000 | 280,000 |
| Fuel cost (€/MWh$_{LHV}$)[6] | 130 | 130 | 100 | 120 | 130 |
| Daily boil-off (%/d) [80] | 0 | 0.1 | 1 | 0 | 0 |
| Travelling speed (kn) | 16 | 16 | 16 | 16 | 16 |
| WACC (%) | 10 | 10 | 10 | 10 | 10 |
| Depreciation period (yr) [70,81] | 27 | 27 | 27 | 27 | 27 |
| CAPEX storage tanks (€/MWh$_{LHV}$) [66,77] | 0.083 | 0.305 | 0.831 | 0.144 | 0.083 |
| CAPEX ship[7] (M€) [own assumptions after 41] | 142.76 | 148.67 | 196.66 | 251.09 | 142.76 |
| OPEX fix (M€/yr) [41] | 3.48 | 3.60 | 7.74 | 5.26 | 3.48 |

All further techno-economic assumptions e.g. to the desalination plants can be found in section B of the Appendix.

## 4. Results

In the following, we first describe the results of the potential area analysis, followed by the results of the cost optimal modelling of PtX fuel production facilities. Subsequently, the production quantities are presented. The chapter closes with the results of the transport cost calculation from selected export countries to Germany and a short portrait of the WebGIS application illustrating the results. All results are integrated into this application and made openly available to the public in the form of the Global PtX Atlas [82].

### 4.1. Area identification

The global analysis of the area identification reveals substantial potential areas of over 32 million km² for the use of onshore wind turbines and/or PV ground-mounted systems. After considering technical and ecological restrictions (cf. subsection 2.1.1), an area of about 2.6 million km² remains for PtX technologies, of which 71 % is attributed to inland waters and 29 % to coastal waters. The distribution of the areas between pure wind sites (38 %), pure PV sites (26 %) or hybrid sites as a combination of wind and PV (36 %) indicates a tendency towards increased use of wind energy. The identified PtX potential areas are distributed over 97 countries, of which 42 countries have relevant potentials bigger than 2,500 km².

An exemplary illustration of the area identification based on the criteria defined in subsection 2.1 is shown in Fig. 6. The exclusion criteria on the left side and the considered areas on the right side are shown for a section of North Africa. In particular, the exclusion due to lack of infrastructure (Fig. 6 B) and water availability (Fig. 6 C) is evident. Looking at the right side of the illustration, large PtX potential areas along coastal water but also next to inland waters, see Egypt next to the Nile River, become obvious.

---

[6] The fuels costs assumptions are based on the worldwide average PtX production costs from the PtX system optimisation from this study [79].
[7] Without the costs for the storage tanks

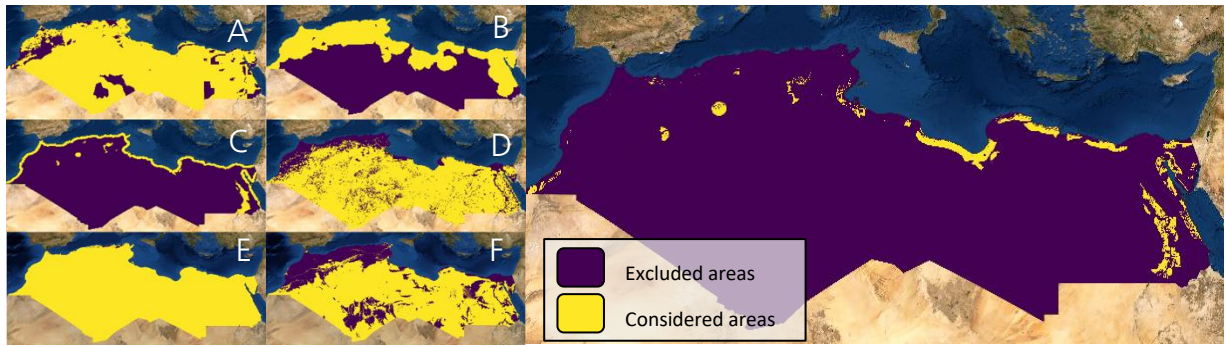

*Fig. 6: Illustration of the area identification using an example region in North Africa. On the left, the grouping of the exclusion criteria to consider nature conservation (A), infrastructure (B), water availability (C), unsuitable areas (D), PV LCOE (E) and Wind LCOE (F). On the right, the resulting areas of interest (Source: Own illustration, basemap from [83]).*

The percentage distribution of global area potential by continent and total area of preferred PtX regions are shown in Fig. 7. Since freshwater is needed for electrolysers, inland waters are attractive sites for PtX, provided they offer good conditions for wind energy and/or PV and do not have water stress. The most significant potentials along inland waters are in the United States, Argentina and Australia. Africa exhibits mostly PtX potentials next to coastal waters.

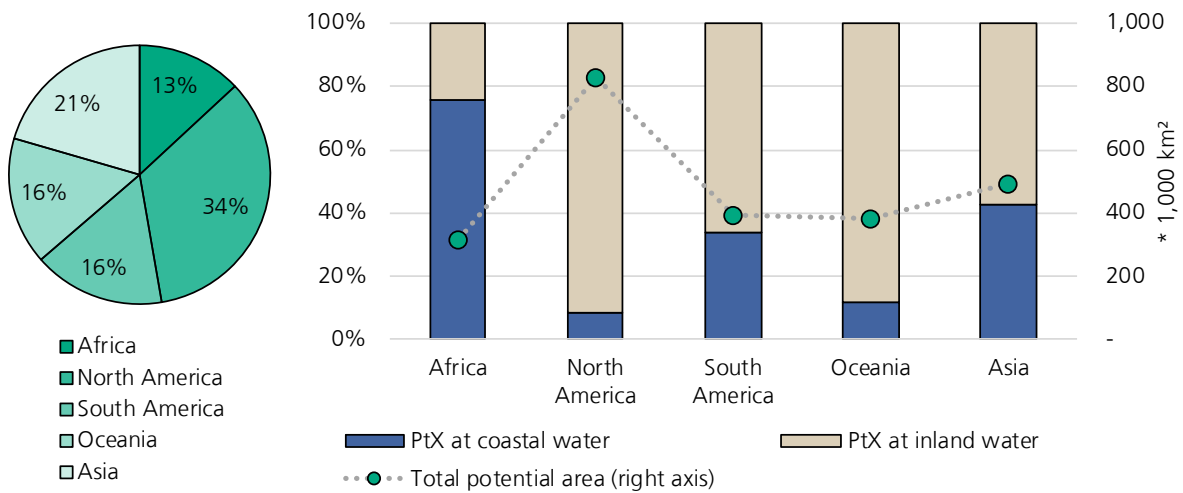

*Fig. 7: Percentage distribution and total area of the preferred PtX regions divided by water supply source (Source: Own illustration).*

Ten countries alone represent 80 % of the globally identified PtX area potential and these countries are presented in Fig. 8. The largest PtX potentials are shown in the United States, followed by Australia, Argentina and Russia[8] (left part of the graph). More specifically, the United States and Australia also show high potential in the socioeconomic analysis. Other countries with high socioeconomic potential and large areas with potential include Canada and Chile. Countries on the African continent, i.e. Egypt, Libya, also exhibit high PtX area potential, but the socioeconomic potential is significantly lower here. Australia has the largest PtX potential at pure PV locations, Russia the largest for pure wind locations and the United states for hybrid locations.

---

[8] We would like to point out that the consequences of the war in Ukraine couldn't be considered, as all steps of the analyses were conducted in 2021 or earlier. The war affects e.g. the area identification and especially the socio-economic evaluation.

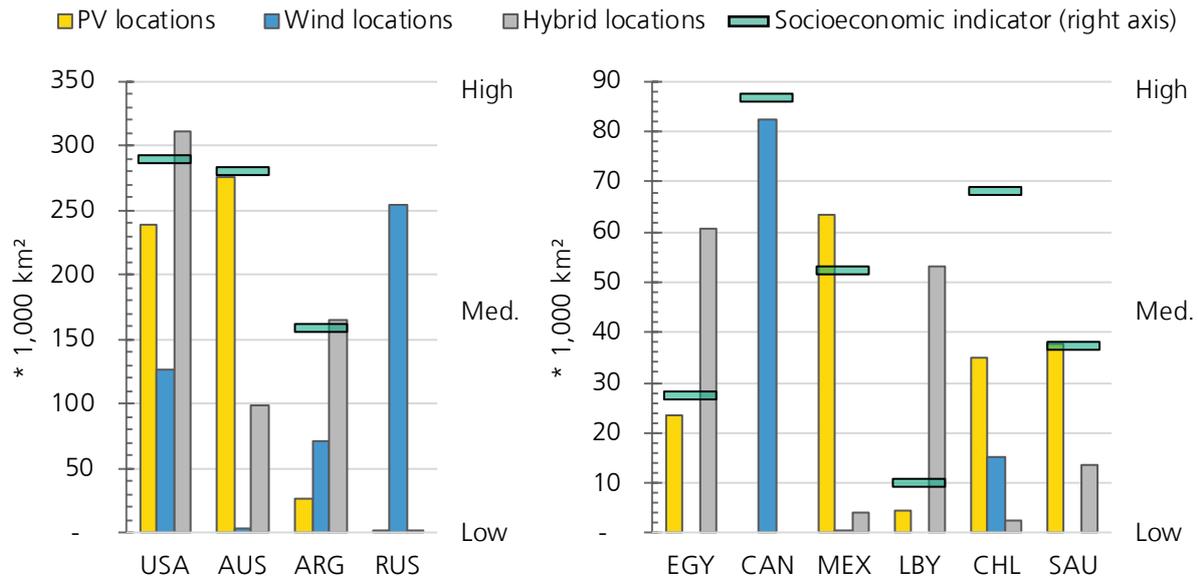

*Fig. 8: Country overview with the ten largest PtX area potentials separated by PV, wind and hybrid locations as well as their socio-economic potential. For Russia, no evaluation of the socioeconomic analysis is feasible due to the current Russo-Ukrainian war (Source: Own illustration).*

### 4.2. System design and fuel production costs

This section gives an overview of the results for the cost-optimal modelling of PtX fuel generation facilities. The results contain, among other aspects, the composition of the various components of the generation plant for each of the 14 production pathways at all of the almost 600 (potential) production sites. Furthermore, the output of the SCOPE model includes hourly resolved time series for almost every component, e.g. the RE sources, the expanded storage technologies, or the output of the corresponding synthesis.

Fig. 9 shows examples of time series for the most important components of a modelled production facility at the coast of Tunisia. The figure displays the first days of the year (January) for a calculation with the historical meteorological year 2008. At the upper graph of Fig. 9, the production of hydrogen and the derived hydrocarbons respectively follow the availability of electricity from the RE sources. Sections of lower electricity production are partly bridged by the available battery storage. At the lower graph, the course of the heat consumption from the DAC unit and the high-temperature electrolyser follows the output schedules of the synthesis and the electrical consumption of the electrolyser, respectively. Periods with high availability of electricity are used to fill up the heat storage (dashed line in the lower graph shows the reservoir level of the heat storage), which is emptied during periods of low electricity availability. During longer periods of lower heat demand, the reservoir level of the heat storage remains at a high level, e.g. see the period from hour 168 to hour 192.

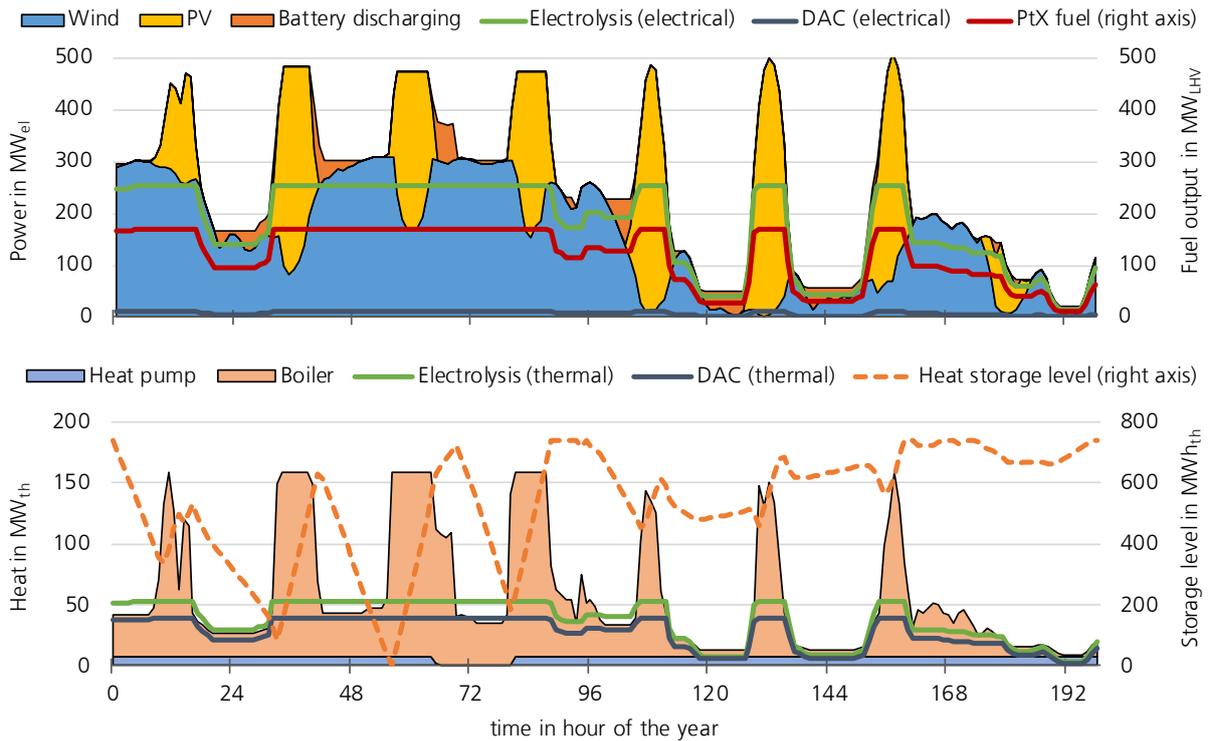

Fig. 9: Extract of a time series of a coastal production site in Tunisia for the production of Fischer-Tropsch fuel with hydrogen from high temperature electrolysis. Upper graph shows the power generation and consumption as well as the fuel output. Lower graph shows the heat generation and consumption (Source: Own illustration).

### 4.2.1. Cost comparison of single fuel production pathways

Generally, the best production sites, which in this context corresponds to the cheapest, are located in Southern America, i.e. Chile, Argentina and Venezuela. These are pure wind sites. The best sites for purely PV-based production sites are in Chile and Peru. The best hybrid sites are located in Mauretania and Venezuela. The comparison between the 14 production pathways and across all simulated production sites is given by a boxplot in Fig. 10. In general, the production costs of ammonia are lower compared to methanol and hydrocarbons due to the high cost of the DAC technology. The cheapest production pathway is the production of gaseous hydrogen via a SOEC. In contrast, the production of liquid methane or Fischer-Tropsch fuels via PEM electrolysis is the most expensive pathway for the best sites. The production costs of the hydrocarbons and ammonia cover a wider range when using SOEC electrolyser than when using PEM electrolyser. Note that the best simulated production sites tend to always be slightly cheaper for the SOEC production pathways, whereas the worst sites are mainly more expensive for the SOEC production pathways. This cost disadvantage of the SOEC is due to its poor load change behaviour and the coupling of the electrolyser with synthesis and DAC/ASU technology (we do not model intermediate storage for $CO_2$ or $N_2$). For these reasons, the goal is to maximize the capacity factor of the SOEC electrolyser. For pure PV or hybrid sites with a very high percentage of PV systems in the RE generation structure, high capacity factors of the electrolyser can be achieved by adding large battery storage. This shifts PV generation peaks into the night. Production pathways with PEM electrolysis instead allow larger electrolyser capacities with lower utilisation rates, which is cheaper than building large battery storages.

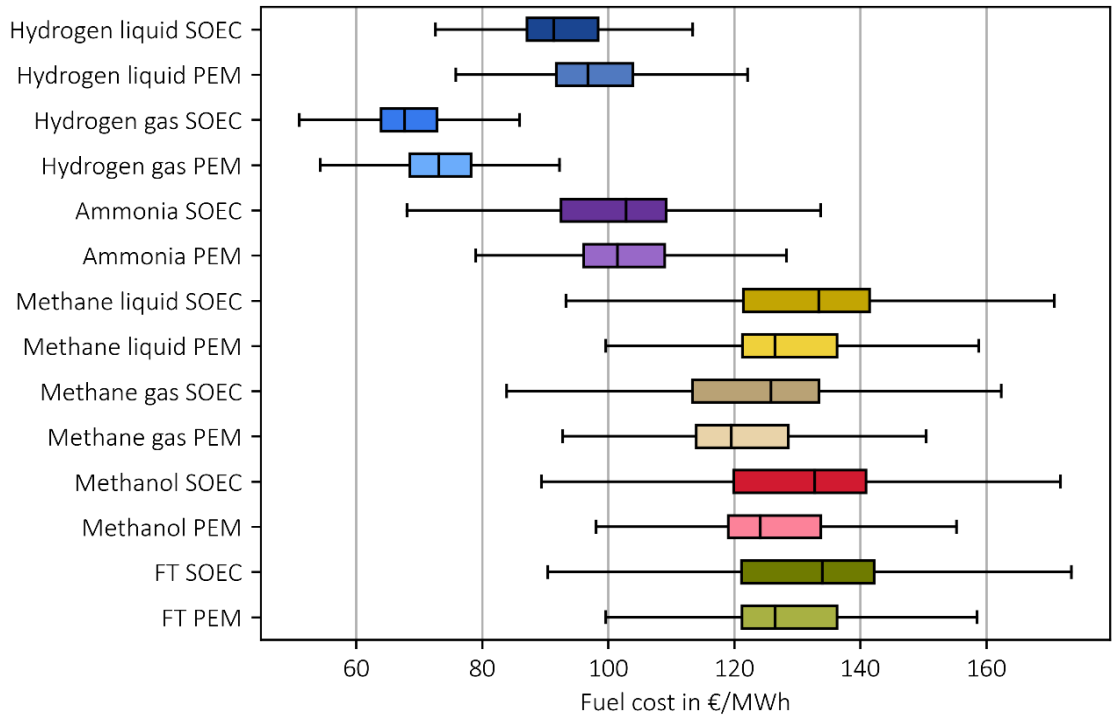

*Fig. 10: Boxplot of the fuel production costs of all simulated production sites displayed per production pathway (combination of fuel and electrolyser technology). Outliers that are more expensive than the upper quartile plus 1.5 of the interquartile range or cheaper than the lower quartile minus 1.5 of the interquartile range are sorted out (Source: Own illustration).*

Looking at selected regions in South America, Australia or parts of Africa and the MENA Region in Fig. 11 the spatial distribution of the production costs in the example of gaseous hydrogen becomes obvious. Best sites for the production are located in the south of Chile and Argentina or in the trade wind regions of Africa in Mauretania and Somalia. South America shows a wide range of production costs, whereas Australia or the MENA region show minor differences.

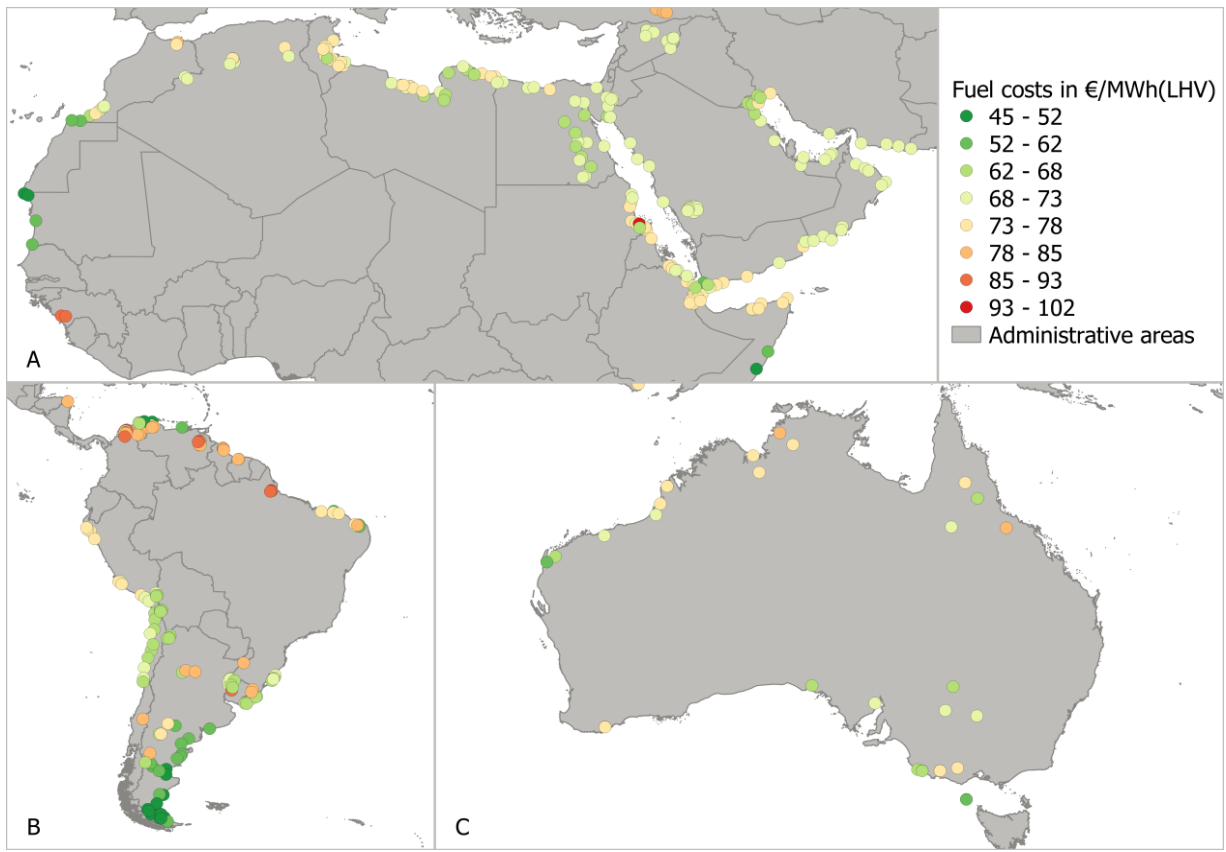

Fig. 11: Illustration of the fuel production costs of gaseous hydrogen via PEM electrolyser in selected regions of Africa and the MENA Region (A), South America (B) or Australia (C). (Source: Own illustration, administrative areas from [43])

To point out the differences between the various categories of production sites we show in the following subsections cost ranges for two fuel types, gaseous hydrogen and Fischer-Tropsch fuel.

### 4.2.2. Gaseous hydrogen

It has already been established that gaseous hydrogen forms the lower limit of the production costs. Among the production options of gaseous hydrogen, the cheapest simulated production sites are pure wind sites at coastal water in Chile with costs of 42.36 €/MWh$_{LHV}$ and pure wind sites at inland water with 42.34 €/MWh$_{LHV}$, respectively (cf. Fig. 12). Pure wind sites also show the largest variation in production costs between the best and the worst production sites. An important reason for this is the large difference in the capacity factor at wind sites (approx. 28.5 % capacity factor at worst sites vs. 68.5 % at best sites) compared to the pure PV sites (12.5 % capacity factor at worst vs. approx. 22.8 % at best sites). At hybrid sites, the two RE sources complement each other, which results in similar low differences in fuel production costs such as those at pure PV sites. The best pure PV sites have fuel costs of 58.70 €/MWh$_{LHV}$ at inland waters and 60.55 €/MWh$_{LHV}$ at coastal waters respectively (both in Chile). The cheapest hybrid sites are in Venezuela with 44.64 €/MWh$_{LHV}$ at coastal waters and, again, in Argentina with production costs of 52.99 €/MWh$_{LHV}$ for sites at inland waters.

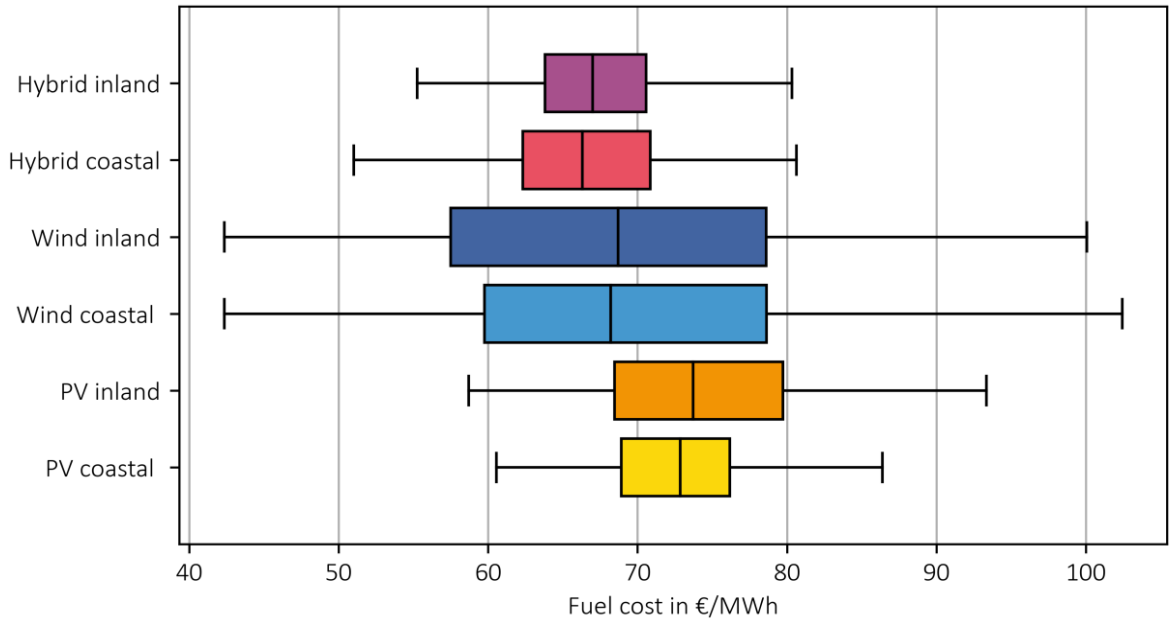

*Fig. 12: Boxplot of the fuel production costs of gaseous hydrogen via SOEC and PEM electrolyser, displayed for the six different site categories (Source: Own illustration).*

### 4.2.3. Fischer-Tropsch fuels

The production costs for the more expensive Fischer-Tropsch fuels are shown in Fig. 13. Due to the additional electricity and heat demand from the DAC unit, the differences between the best and the worst sites are even larger than to produce hydrogen (subsection 4.2.2).

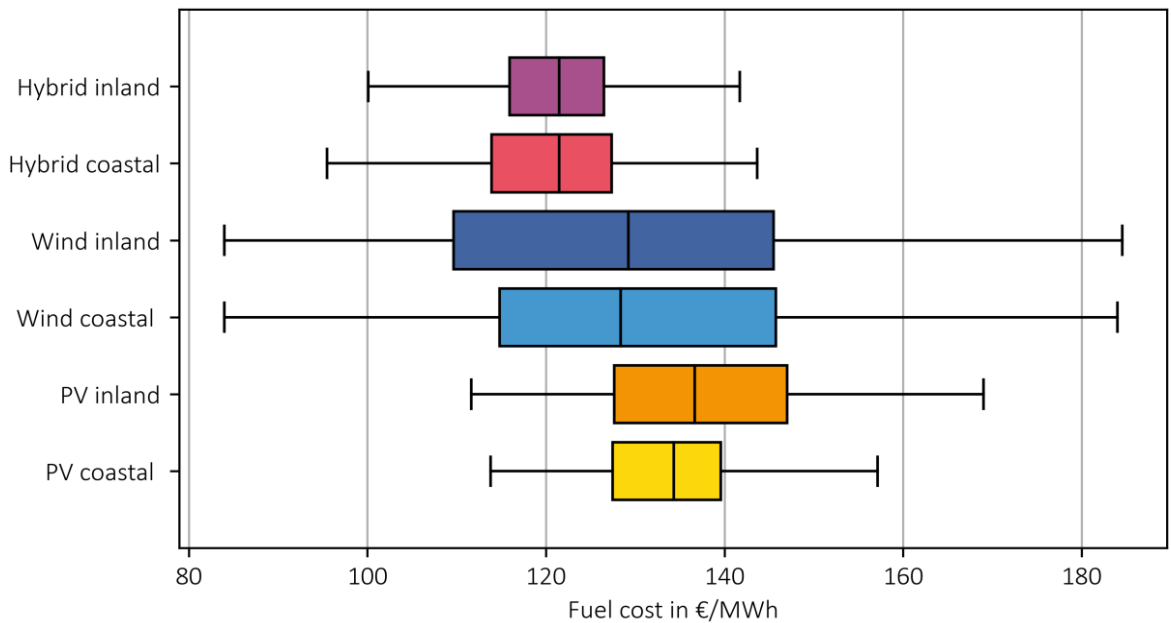

*Fig. 13: Boxplot of the fuel production costs of Fischer-Tropsch fuels via SOEC and PEM electrolyser, displayed for the six different site categories (Source: Own illustration).*

The best sites to produce Fischer-Tropsch fuels remain the same as for gaseous hydrogen due to the good weather conditions. The lowest production costs of Fischer-Tropsch fuels are 84.00 €/MWh$_{LHV}$ and 83.97 €/MWh$_{LHV}$ for coastal and inland sites, respectively, and are represented by pure wind sites in Chile. The production costs at pure PV sites also do not differ considerably between coastal (113.81 €/MWh$_{LHV}$) and inland sites (111.63 €/MWh$_{LHV}$). However, at hybrid sites, the differences are more pronounced between coastal (88.21 €/MWh$_{LHV}$) and inland sites (100.11 €/MWh$_{LHV}$). Looking at the average costs, the hybrid locations show the best. Moreover, the range is most robust here. The widest range is shown by the wind categories and the highest average cost can be found in the PV categories.

### 4.3. Production quantity potentials
#### 4.3.1. Overview

According to subsection 2.4, the PtX fuel production quantity is calculated for every country and every PtX pathway. The aggregated production quantity can be seen in Fig. 14.

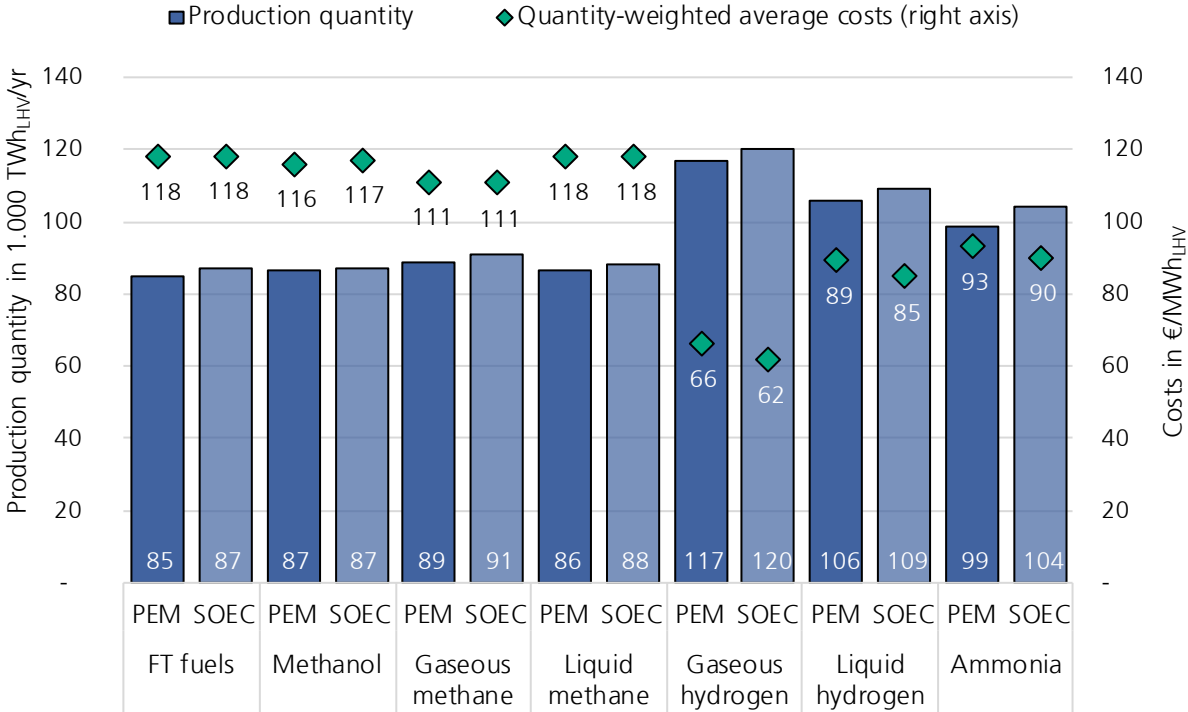

Fig. 14: Illustration of the production quantity and quantity-weighted average production costs of all estimated PtX pathways that can be derived from the potential areas and the optimisation of specific sites in each country (Source: Own illustration).

The cumulative production quantity for gaseous hydrogen sums up to almost 120,000 TWh/yr (using high temperature SOEC electrolysis), whereas the value for hydrocarbons ranges between 85,000 TWh/yr and 88,000 TWh/yr, depending on the production pathways. Taking socioeconomic aspects (subsection 2.6) into account, the potential is reduced by 37 % to then 75,600 TWh/yr of hydrogen or 54,800 TWh/yr respectively 53,550 TWh/yr of hydrocarbons.

Four of the five analysed continents show similar potential production quantity of approx. 16,840 TWh/yr in Australia up to almost 22,622 TWh/yr in Asia. With a possible production quantity of more than 39,248 TWh/yr, North America has approximately twice the potential of the other single continents. With the exception of Africa, the potential of inland sites is greater than that of coastal sites on all other continents. In Australia, pure PV sites are the most relevant RE source category, whereas in Asia (mainly dominated by Russia) pure wind sites are predominant. For the other continents, hybrid sites are the determining category of RE source.

The results aggregated for each continent and separated by coastal and inland sites to produce gaseous hydrogen in combination with PEM electrolysis can be seen in Fig. 15, and is exemplary for all pathways.

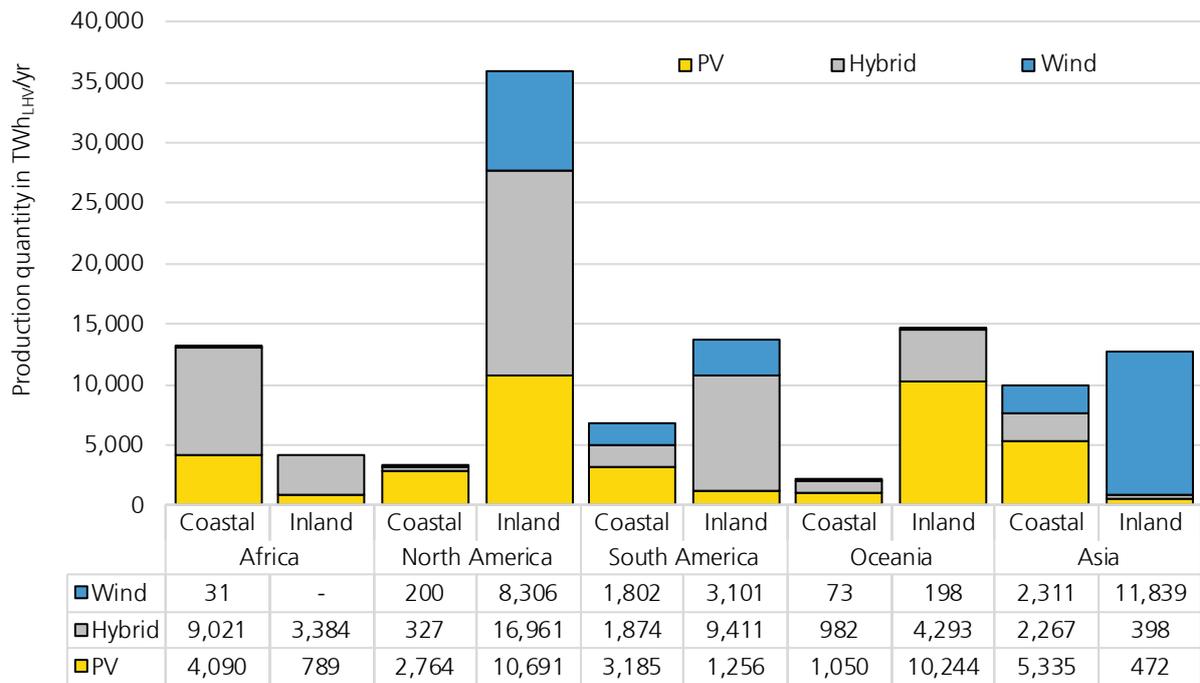

*Fig. 15: Illustration of the production quantity of gaseous hydrogen via PEM electrolyser that can be derived from the potential areas and the optimisation of specific sites in each country. The results are aggregated by continents and water source and separated by each combination of RE (Source: Own illustration).*

To point out the differences between the production sites we illustrate in the following subsections the production quantity in combination with a cost function for two fuel types, gaseous hydrogen and Fischer-Tropsch fuel.

### 4.3.2. Gaseous hydrogen

Fig. 16 displays the production cost curve of gaseous hydrogen using a PEM electrolyser in 2050. In addition, the related RE source is highlighted in the chart. The most cost-efficient sites fall below the mark of 51 €/MWh$_{LHV}$, but the corresponding production quantity is limited at 3,000 TWh/yr. Between 51 and 61 €/MWh$_{LHV}$ the production quantity increases due to equally available wind and hybrid sites and up to thirty percent of the total production quantity. Only a few PV sites can compete with costs at around 64 €/MWh$_{LHV}$. Compared to wind (about 22,000 TWh/yr) and hybrid sites (about 35,000 TWh/yr) only approx. 2,400 TWh/yr can be produced below 70 €/MWh$_{LHV}$ at PV sites. 73 % of the potential production quantity of hybrid sites and 78 % for wind sites are below 70 €/MWh$_{LHV}$. Although PV sites account for one third of the whole production quantity, only around 6 % of this quantity is cheaper than 70 €/MWh and thus competitive to wind and hybrid production sites.

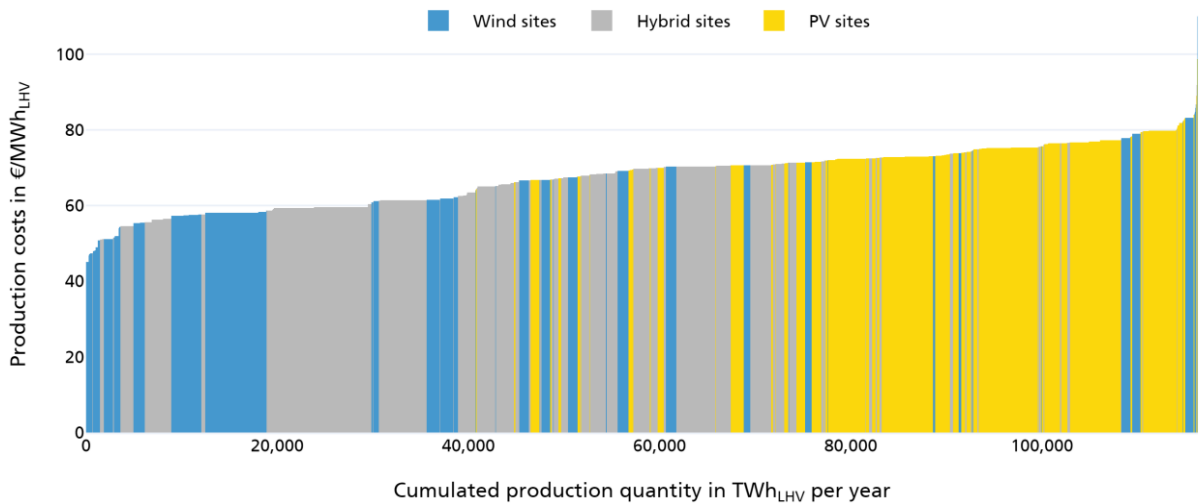

Fig. 16: Illustration of the production cost curve of gaseous hydrogen using a PEM electrolyser in 2050. The potential production quantity is divided into PV (yellow area), wind (blue area) and hybrid sites (grey area). The quantity is shown for each of the simulated representative site types and sorted from the cheapest to the most expensive production costs. (Source: Own illustration).

### 4.3.3. Fischer-Tropsch fuels

The following illustration shows the production quantity and the production costs for Fischer-Tropsch-fuels (Fig. 17). Only small amounts of Fischer-Tropsch fuels at wind and hybrid sites can be produced at a cost of less than 90 €/MWh$_{LHV}$. The results show that 26,500 TWh/yr are possible at costs below 113 €/MWh$_{LHV}$. For this part of the supply, only wind and hybrid plants are relevant. In contrast to gaseous hydrogen, there are also relevant amounts of wind (26 %) and hybrid quantities (16 %) available in more expensive ranges at costs beyond 126 €/MWh$_{LHV}$. Overall, PV sites are more competitive for the production of Fischer-Tropsch fuels compared to wind and hybrid sites than for the production of gaseous hydrogen.

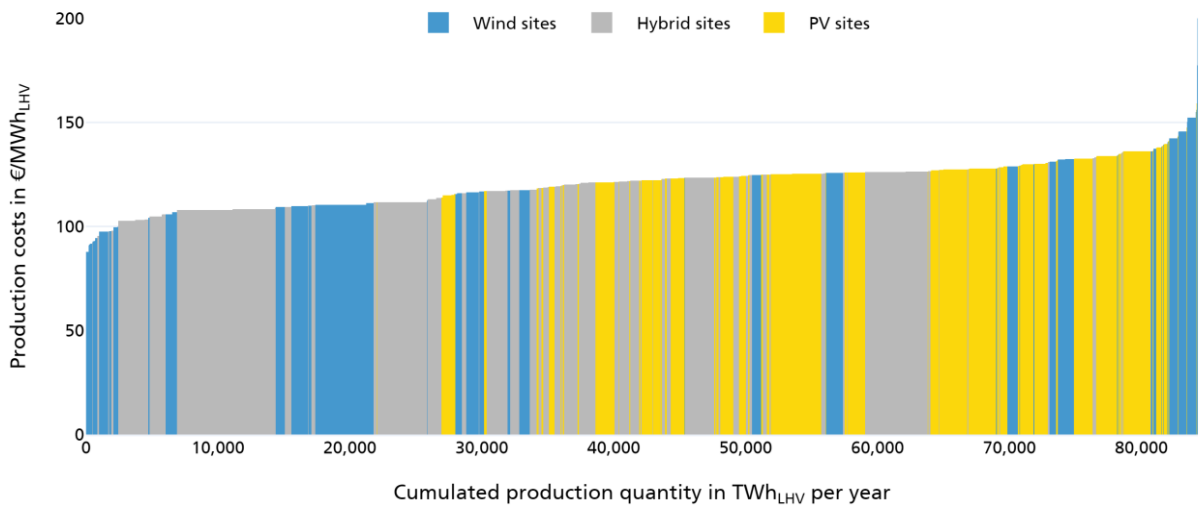

Fig. 17: Illustration of the production cost curve of Fischer-Tropsch using a PEM electrolyser in 2050. The potential production quantity is divided into PV (yellow area), wind (blue area) and hybrid sites (grey area). The quantity is shown for each of the simulated representative site types and sorted from the cheapest to the most expensive production costs. (Source: Own illustration).

### 4.4. Transport costs to Germany

This section shows the results of the calculation of the transport costs (cf. 2.5) as an example for all liquid energy carriers considered such as Fischer-Tropsch fuels, methanol, liquid methane, liquid hydrogen and ammonia from selected countries to Germany. Selected are the countries with a medium or high socioeconomic potential and a PtX production quantity potential of at least 100 TWh$_{LHV}$/yr. Further transport costs from the production countries to all countries in the European Union can be viewed online at the Global PtX Atlas.

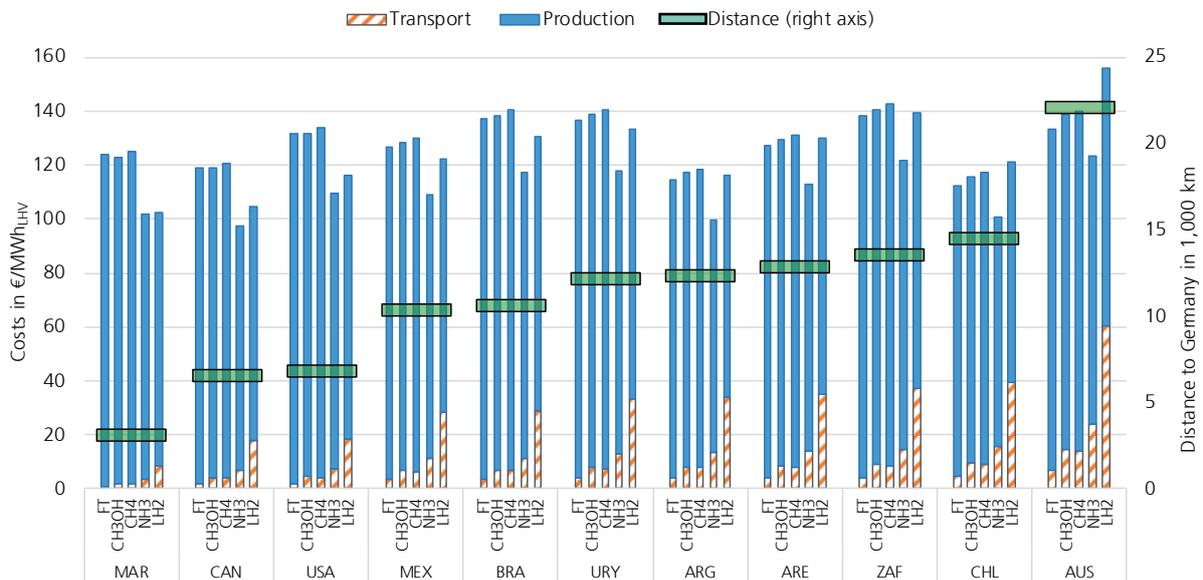

*Fig. 18: Average production costs in the producing country and transport costs to Germany for Fischer-Tropsch fuels (FT), methanol (CH3OH), liquid methane (CH4), liquid ammonia (NH3) and liquid hydrogen (LH2) via PEM electrolyser. Countries are sorted from lower to higher transport distance to Germany (MAR = Morocco, CAN = Canada, USA = United States of America, MEX = Mexico, BRA = Brazil, URY = Uruguay, ARG = Argentina, ARE = United Arab Emirates, ZAF = South Africa, CHL = Chile, AUS = Australia; Source: Own illustration).*

Considering the transport by ship to Germany, there are significant differences in the PtX fuels (s. Fig. 18). When comparing the mentioned options, ammonia shows the lower limit of import costs to Germany from all selected countries. Ammonia transport from nearby regions (e.g. Morocco) shows a cost saving of about 18 % compared to Fischer-Tropsch fuels. A cost advantage is also shown from distant regions such as Australia, although much reduced at 7 %. In comparison, liquid hydrogen is not competitive. The transport of liquid hydrogen is energy-intensive (due to the boil-off losses of hydrogen) and thus expensive. In Australia, for example, hydrogen can be produced cost-effectively but the long transport distance makes this location one of the most expensive. On the other hand, Morocco, which is relatively close to Germany, shows competitive import costs of liquid hydrogen relative to ammonia. In countries such as South Africa or Uruguay, the cost advantages of hydrogen production are reduced due to the transport costs, leading to almost identical import costs of liquid hydrogen and Fischer-Tropsch fuels. The lower limit of import costs for these countries is shown for Canada with 97 €/MWh$_{LHV}$ and ammonia as fuel.

### 4.5. The Global PtX Atlas

Within the project, a WebGIS application (https://maps.iee.fraunhofer.de/ptx-atlas/, s. Fig. 19) is developed to provide interested users with an interactive and all-encompassing view of the results. Via map function the users get insight into high-resolution GIS analyses for PtX area identification. Based on this, simulation results on generation quantities and future costs can be queried graphically with the help of a sidebar. Within the sidebar, a distinction is made between aggregated and site-specific evaluations.

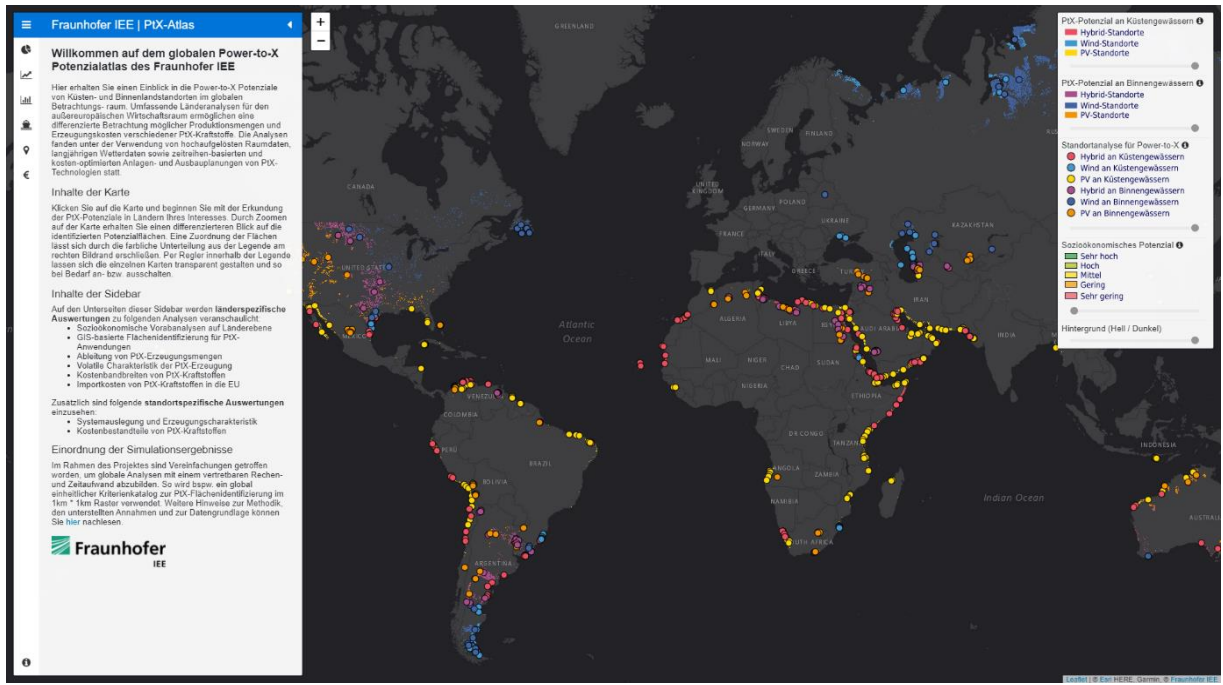

*Fig. 19: Screenshot of the developed WebGIS tool.*

The aggregated evaluations include

- the high-level socioeconomic analysis,
- an aggregation of the GIS-based area identification for PtX applications in the respective country or region,
- the theoretical maximum amount of possible generation from the identified areas,
- the average capacity factor and the volatile generation characteristics per month,
- cost ranges of PtX fuels for all investigated countries and regions and
- import costs to the countries of the European Union.

In addition, site-specific evaluations can also be viewed. These include

- the system design and the capacity factor of the most important system components of all 14 modelled PtX generation systems,
- generation characteristics, using weekly plots for a complete historical weather year with the focus on the generation output of renewables as well as electrolysis and fuel generation and
- cost components of a PtX fuel generation at the site.

## 5. Discussion

During the analysis, simplifications have been made in order to map global potentials with a justifiable computational and time effort. A globally uniform criteria catalogue for PtX area identification is used. The grid resolution is about 1 km × 1 km at the equator, which can not replace detailed on-site analyses. The mapping of renewable energy power generation is based on mesoscale weather models with a temporal resolution of one hour. Especially a site assessment for wind energy usually requires an elaborate wind measurement campaign. However, it allows the identification of preferred sites and provides a first estimation of future potentials at the global level.

Strict criteria for nearby infrastructure availability like water sources, ports, pipelines and cities identify best-located regions. As a result, in some regions the potential areas are enormously restricted, e.g. isolated regions in the interior of the country. For future studies, we need to reflect on whether the exclusion criteria need to be adjusted, or whether these isolated regions should still be designated as potential.

A detailed downstream analysis requires site-specific criteria related to local conditions to consider all necessary influencing factors for a PtX site suitability. Such a site assessment also includes planning the necessary wind and PV plants based on the prevailing weather conditions. Typically, these processes take several years due to their complexity. These additional investigations may result in a significantly lower potential than analysed.

The analysis is based on historical weather data from 2008-2012 for the future year 2050. We expect effects of climate change on the local weather conditions and potentially also on some area identification parameters (e.g. water stress level). Future research should also focus on such influences on the determined PtX potentials.

The cost calculation is based on techno-economic parameters, which are projected for the year 2050 based on an extensive literature research. Since the literature mostly provides wide ranges for future technology developments, medium assumptions were taken from the literature. The calculated costs depend strongly on the techno-economic assumptions and a more optimistic development results in lower costs or a less favourable development results in higher costs. Due to the lack of variation in parameter selection, sensitivity at individual sites is not mapped. Since a large number of locations are simulated, which have a wide range of costs, a comprehensive overview of the future production costs of PtX fuels is nevertheless obtained. In addition, no country-specific differences are taken into account, but a global market was assumed for the long term. This applies to all techno-economic data such as investment and operating costs as well as interest rates.

Furthermore, the optimisation itself is subject to several limitations and simplifications. For example, the partial load behaviour of the individual plant components is not modelled but only represented in a simplified way. Since large-scale plants are modelled, a stack design of the electrolysers is assumed, which are used according to demand. For the technical availability of the plant components, only general reductions are assumed, which depend on renewable resources. The technical availability of all components must be considered independent of each other in order to account for periods of repair and maintenance work. Higher outage times due to maintenance work would increase PtX generation costs.

The modelling of DAC units requires special consideration. DAC is a very young technology in the pilot stage. Many technical questions about efficiency and lifetime are still open. In particular, only limited research has been done on the applicability of the systems in different regions of the world with different climatic conditions. For example, in hot dry desert regions, the efficiency and lifetime of the plants may suffer. One possibility is to consider the impact based on weather parameters such as temperature and relative humidity.

The estimation of the total PtX generation quantity assumes a full theoretical development of the identified areas. It should be noted, however, that the feasibility of market ramp-up is an essential criterion for estimating future generation quantities and costs of PtX products, both for the expansion of renewable energies and for PtX technologies. In addition, the use of RE power for PtX competes with the decarbonisation of local power generation. Therefore, it is necessary to investigate which areas of the identified PtX regions will be devoted to the production of PtX products in the long term.

Modelling of transportation options is reduced to transportation by ship. As a simplification, we always choose the largest port of each country to define the transportation route between the exporting country and the importing country. In order to quantify the transport costs for certain routes more precisely, further ports of the country should be included. Other transport options, such as gaseous transport by pipeline from nearby regions, must also be considered.

Concerning import options to Europe several aspects has to be discussed. The production potential is particularly large in countries like Australia and the United States, which offer excellent spatial and meteorological conditions for producing substantial quantities of PtX fuels. They are also politically stable and offer a reliable investment framework. However, for the USA, it remains to be seen what share of the production quantities will be available for export. Countries in closer proximity to Europe, e.g. Egypt or Libya, would also be able to supply large quantities of PtX fuels, since the transport distances are comparatively short, so that delivery by pipeline is also possible. However, socioeconomic indicators are lower in these countries. Therefore, investment risks are higher, increasing financing costs and reducing the likelihood of large-scale PtX projects being realized there. In addition, in PtX export regions, the use of RE power for PtX competes with the decarbonisation of local power generation. Consideration of national requirements is, of course, necessary for all countries before an export potential can be determined.

Future research should examine the development of the global trade volume and market prices of the respective PtX markets based on multi-criteria approaches, transformation scenarios and detailed modelling of production and transport potentials, as well as demand volumes. It will be valuable to assess the impact of constraints, such as limited trade between democratic and non-democratic nations or limited renewable/electrolyser capacities until 2030, and of synergies such as welfare for cooperating regions on different time scales (medium and long term). When investigating future market situations, the influence of strategic behaviour should also be taken into account with appropriate methods.

## 6. Summary and conclusion

The main goal of the paper was to analyse which regions are capable of producing significant amounts of PtX fuels and at what cost fuel production can occur. We used spatial data with a high resolution to identify worldwide potential PtX production areas. We simulated hourly resolved RE generation time series for the largest areas of every category in each country. With the help of these time series and techno-economic assumptions, we modelled cost optimal system designs of PtX production facilities for 14 different production paths at almost 600 sites. From the results of the potential area identification and the PtX system design, we derived potential production quantities for every fuel. Additionally, we calculated the transport costs from the largest port of the producing countries to Germany for the liquid energy carriers via ship.

The PtX production quantity outside Europe is up to 120,000 TWh/yr of hydrogen or 87,000 TWh/yr of electricity-based liquid fuels in the long term. Taking socioeconomic aspects (subsection 2.6) into account, the potential is reduced by 37 % to then 75,600 TWh/yr of hydrogen or 54,800 TWh/yr of hydrocarbons. The majority (70 %) of these quantities are inland sites and only 30 % near the coast. Large territorial states such as the United States, Australia and Argentina account for 50 % of the identified potentials. Non-European locations offer a lot of potential for the production of PtX fuels - a potential large enough to cover the remaining demand for these energy carriers globally if energy efficiency and direct electricity are preferred.

The costs show a wide range across the simulated locations. The lowest production costs for PtX generation are in Chile, Argentina, Venezuela and in Mauritania with a lower limit of 42.3 €/MWh$_{LHV}$ to 46.5 €/MWh$_{LHV}$ for gaseous hydrogen and 84.0 €/MWh$_{LHV}$ to 89.1 €/MWh$_{LHV}$ for Fischer-Tropsch fuels. These low cost locations mainly represent pure wind locations, but hybrid locations are only marginally more expensive. Looking at the average costs, the PV locations represent the most expensive category. The bandwidth of the production costs within the different categories is diverse. Whereas pure wind locations show a wide spectrum between the cheapest and most expensive sites (cf. Fig. 11 and Fig. 13), these differences are way less significant for pure PV locations. Hybrid locations have the smallest range of production costs between the best and the worst locations, due to the complementation of the electricity generation from wind and PV.

If the transport costs for liquid fuels are taken into account, the assessment of different PtX fuels partly changes. High transport costs of liquid hydrogen decreases cost advantages of the hydrogen paths. For long distances (e. g. Australia to Germany), it is uneconomical to transport green hydrogen and instead liquid hydrocarbons or ammonia are preferable. Canada as a country with a very high socioeconomic indicator shows the lower limit with 97.0 €/MWh$_{LHV}$ and ammonia as fuel.

In conclusion, there is enormous potential for the production of PtX fuels worldwide, although costs differ. While the global RE potential is very large, first PtX project realisations have only occurred in a few regions. Hence, the limiting factor for the expansion of PtX is not the availability of land but rather the maximum possible expansion dynamics for renewable energies and PtX technologies.

## Acknowledgement

This work has been developed in the projects DeV-KopSys and DeV-KopSys-2. DeV-KopSys (reference number: 14EM4005-1) was funded by the German Federal Ministry for the Environment, Nature

Conservation and Nuclear Safety (BMU) and DeV-KopSys-2 (reference number: 16EM5008-1) is funded by the German Federal Ministry for Economic Affairs and Climate Action (BMWK) within the research programme "Renewable Mobile".

We thank Norman Gerhardt, Carsten Pape, Jochen Bard, Daniel Horst and Philipp Härtel for valuable comments and discussions.

## Data Availability

All results of the investigation are freely available via a developed WebGIS application. This application is called Global PtX Atlas and is accessible on https://maps.iee.fraunhofer.de/ptx-atlas/.

## A Appendix SCOPE SD model

In the following the SCOPE SD model for calculating the cost optimal system configuration for the generation of synthetic fuels is explained. Section A. 1 lists all used variables and section A. 2 shows the used parameters. Section A. 3 contains the model equations that represent the technical system while the goal function of the optimisation model is presented in section A. 4.

The model has an hourly time resolution and the index $t \in T = \{1, \ldots, 8760\}$ is used for the time steps. All variables and parameters are additionally associated to the analysed single sites with different local weather conditions. For reasons of a better readability there is no index used for the sites nor for the 14 different technology variants.

### A. 1 Variables

All variables used in the SCOPE SD model are non-negative.

| Symbol | Description |
|---|---|
| $fuel_t^{(\cdot),final}$ | Fuel output of synthesis (e.g. $CH_4$, $CH_3OH$,…) in MWh$_{fuel,\,LHV}$ |
| $fuel_t^{CH_4,liquefy}$ | $CH_4$ input of liquefaction in MWh$_{CH4,\,LHV}$ |
| $fuel_t^{H_2,liquefy}$ | $H_2$ input of liquefaction in MWh$_{H2,\,LHV}$ |
| $fuel_t^{H_2,synthesis}$ | $H_2$ input of synthesis in MWh$_{H2,\,LHV}$ |
| $p_t^{battery,con,LC+/-}$ | Load change of battery storage power consumption in MWh$_{el}$/h |
| $p_t^{battery,gen,LC+/-}$ | Load change of battery storage power generation in MWh$_{el}$/h |
| $p_t^{con,battery}$ | Power consumption from battery storage in MWh$_{el}$/h |
| $p_t^{con,el.boiler}$ | Power consumption of electric boiler in MWh$_{el}$/h |
| $p_t^{con,electrolysis}$ | Power consumption of electrolysis in MWh$_{el}$/h |
| $p_t^{con,processing}$ | Power consumption of further processing steps (liquefaction/compression) of $H_2$ or $CH_4$ in MWh$_{el}$/h |
| $p_t^{con,dac}$ | Power consumption of direct air capture in MWh$_{el}$/h |
| $p_t^{electrolysis,LC+/-}$ | Load change of electrolysis in MWh$_{el}$/h |
| $p_t^{gen,battery}$ | Power generation from battery storage in MWh$_{el}$/h |
| $p_t^{gen,pv}$ | Power generation from solar photovolatics in MWh$_{el}$/h |
| $p_t^{gen,wind_i}$ | Power generation from wind turbines (type $i \in \{1,2\}$) in MWh$_{el}$/h |
| $P^{battery}$ | Installed power of battery storage in MW$_{el}$ |
| $P^{dac}$ | Installed capacity of direct air capture in MW$_{el}$ |
| $P^{el.boiler}$ | Installed capacity of electric boiler in MW$_{el}$ |
| $P^{electrolysis}$ | Installed capacity of electrolysis in MW$_{el}$ |
| $P^{heat\ pump}$ | Installed capacity of heat pump in MW$_{el}$ |
| $P^{processing}$ | Installed capacity of further processing steps (liquefaction/compression) of $H_2$ or $CH_4$ in MW$_{el}$ |
| $P^{pv}$ | Installed capacity of solar photovoltaics in MW$_{el}$ |
| $P^{synthesis}$ | Installed capacity of synthesis in MW$_{H2}$ |
| $P^{wind_i}$ | Installed capacity of wind turbines (type $i \in \{1,2\}$) in MW$_{el}$ |
| $q_t^{el.boiler,electrolysis}$ | Heat of electric boiler to be used by the electrolysis in MWh$_{th}$/h |
| $q_t^{el.boiler,th.storage}$ | Heat of electric boiler to be stored in the thermal storage in MWh$_{th}$/h |

| Symbol | Description |
|---|---|
| $q_t^{heat\ pump,th.storage}$ | Heat of electric heat pump to be stored in the thermal storage in MWh$_{th}$/h |
| $q_t^{synthesis,electrolysis}$ | Waste heat of the synthesis to be used by the electrolysis in MWh$_{th}$/h |
| $soc_t^{battery}$ | State of charge of battery storage in MWh$_{el}$ |
| $soc_t^{CH_4}$ | State of charge of CH$_4$ storage in MWh$_{CH4,\ LHV}$ |
| $soc_t^{H_2}$ | State of charge of H$_2$ storage in MWh$_{H2,\ LHV}$ |
| $soc_t^{Q}$ | State of charge of thermal storage in MWh$_{th}$ |
| $SOC^{battery}$ | Installed capacity of battery storage in MWh$_{el}$ |
| $SOC^{CH_4}$ | Installed capacity of CH$_4$ storage in MWh$_{CH4,\ LHV}$ |
| $SOC^{H_2}$ | Installed capacity of H$_2$ storage in MWh$_{H2,\ LHV}$ |
| $SOC^{Q}$ | Installed capacity of thermal storage in MWh$_{th}$ |

## A. 2 Parameters

All parameters used in the SCOPE model are listed below.

| Symbol | Description |
|---|---|
| $\alpha^{battery}$ | Power-to-capacity ratio of battery storage in MW$_{el}$/MWh$_{el}$ |
| $\beta^{battery}$ | Charging-to-discharging power ratio of battery storage in MW$_{el}$/MW$_{el}$ |
| $\gamma_t^{humidity}$ | Factor describing the ambient humidity which influences the DAC efficiency in % |
| $\delta^{heat,dac}$ | Heat demand factor of DAC in MWh$_{th}$/MWh$_{CHx,\ LHV}$ |
| $\delta^{heat,electrolysis}$ | Heat demand factor of electrolysis in MWh$_{th}$/MWh$_{H2,\ LHV}$ |
| $\eta^{battery,con}$ | Efficiency of battery storage charging (power consumption) in MWh$_{el}$/MWh$_{el}$ |
| $\eta^{battery,gen}$ | Efficiency of battery storage discharging (power generation) in MWh$_{el}$/MWh$_{el}$ |
| $\eta^{el.boiler}$ | Efficiency of electric boiler in MWh$_{th}$/MWh$_{el}$ |
| $\eta^{electrolysis}$ | Efficiency of electrolysis in MWh$_{el}$/MWh$_{fuel,\ LHV}$ |
| $\eta^{heat\ pump}$ | Efficiency of heat pump (or coefficient of performance, COP) in MWh$_{th}$/MWh$_{el}$ |
| $\eta^{processing}$ | Efficiency of further processing steps (liquefaction/compression) of H$_2$ or CH$_4$ in MWh$_{el}$/MWh$_{fuel,\ LHV}$ |
| $\eta^{synthesis}$ | Efficiency of synthesis in MWh$_{fuel,\ LHV}$/MWh$_{H2,\ LHV}$ |
| $l^{battery}$ | Loss of battery storage charging in %/h |
| $l^{H2\ storage}$ | Loss of H$_2$ storage charging |
| $\rho^{heat} \in \{0,1\}$ | Binary indicator which 1 = technology with heat demand and 0 = technology without heat demand |
| $\rho^{H2\ storage} \in \{0,1\}$ | Binary indicator which 1 = system with hydrogen storage and 0 = system without hydrogen storage |
| $\rho^{synthesis} \in \{0,1\}$ | Binary indicator which 1 = system with synthesis process and 0 = system without synthesis process |
| $\rho^{CH4\ storage} \in \{0,1\}$ | Binary indicator which 1 = system with methane storage and 0 = system without methane storage |
| $\rho^{processing} \in \{0,1\}$ | Binary indicator which 1 = technology with further processing (liquefaction/compression) of H$_2$ or CH$_4$ and 0 = technology without further processing (liquefaction/compression) of H$_2$ or CH$_4$ |
| $\theta^{synthesis}$ | Factor for the waste heat of the synthesis process that can be used by the electrolysis in MWh$_{th}$/MWh$_{H2,\ LHV}$ |
| $\bar{\sigma}^{waste\ heat}$ | Maximum share of electrolysis heat demand that may be supplied by synthesis waste heat |
| $\tau^{electrolysis}$ | Stand by heat demand factor of electrolysis in % |
| $AV_t^{(\cdot)} \in [0,1]$ | Hourly availability of solar photovoltaic or wind power (demanding on local weather conditions) |
| $C^{fix,battery}$ | Fixed operating costs of battery storage in €/MW$_{el}$ |
| $C^{fix,CH4\ storage}$ | Fixed operating costs of CH$_4$ storage in €/MWh$_{CH4,\ LHV}$ |
| $C^{fix,dac}$ | Fixed operating costs of direct air capture in €/MW$_{el}$ |
| $C^{fix,el.boiler}$ | Fixed operating costs of electric boiler in €/MW$_{el}$ |
| $C^{fix,electrolysis}$ | Fixed operating costs of electrolysis in €/MW$_{el}$ |
| $C^{fix,H2\ storage}$ | Fixed operating costs of H$_2$ storage in €/MWh$_{H2,\ LHV}$ |
| $C^{fix,heat\ pump}$ | Fixed operating costs of heat pump in €/MW$_{el}$ |
| $C^{fix,synthesis}$ | Fixed operating costs of synthesis in €/MWh$_{H2,\ LHV}$ |
| $C^{fix,pv}$ | Fixed operating costs of solar photovoltaics in €/MW$_{el}$ |
| $C^{fix,th.storage}$ | Fixed operating costs of thermal storage in €/MWh$_{th}$ |
| $C^{fix,wind_i}$ | Fixed operating costs of wind turbines (type $i \in \{1,2\}$) in €/MW$_{el}$ |
| $C^{inv,battery}$ | Equivalent annual investment costs of battery storage in €/MW$_{el}$ |
| $C^{inv,CH4\ storage}$ | Equivalent annual investment costs of CH$_4$ storage in €/MWh$_{CH4,\ LHV}$ |
| $C^{inv,dac}$ | Equivalent annual investment costs of direct air capture in €/MW$_{el}$ |
| $C^{inv,el.boiler}$ | Equivalent annual investment costs of electric boiler in €/MW$_{el}$ |
| $C^{inv,electrolysis}$ | Equivalent annual investment costs of electrolysis in €/MW$_{el}$ |
| $C^{inv,H2\ storage}$ | Equivalent annual investment costs of H$_2$ storage in €/MWh$_{H2,\ LHV}$ |
| $C^{inv,heat\ pump}$ | Equivalent annual investment costs of heat pump in €/MW$_{el}$ |
| $C^{inv,synthesis}$ | Equivalent annual investment costs of synthesis in €/MWh$_{H2,\ LHV}$ |
| $C^{inv,pv}$ | Equivalent annual investment costs of solar photovoltaics in €/MW$_{el}$ |
| $C^{inv,th.storage}$ | Equivalent annual investment costs of thermal storage in €/MWh$_{th}$ |
| $C^{inv,wind_i}$ | Equivalent annual investment costs of wind turbines (type $i \in \{1,2\}$) in €/MW$_{el}$ |
| $C^{LC,battery}$ | Load change costs of battery storage in €/MW$_{el}$ |
| $C^{LC,electrolysis}$ | Load change costs of electrolysis in €/MW$_{el}$ |
| $C^{var,battery}$ | Variable operating costs of battery storage in €/MWh$_{el}$ |
| $C^{var,CH4\ storage}$ | Variable operating costs of CH$_4$ storage in €/MWh$_{CH4,\ LHV}$ |
| $C^{var,dac}$ | Variable operating costs of direct air capture in €/MWh$_{el}$ |
| $C^{var,el.boiler}$ | Variable operating costs of electric boiler in €/MWh$_{el}$ |
| $C^{var,electrolysis}$ | Variable operating costs of electrolysis in €/MWh$_{el}$ |
| $C^{var,H2\ storage}$ | Variable operating costs of H$_2$ storage in €/MWh$_{H2,\ LHV}$ |
| $C^{var,heat\ pump}$ | Variable operating costs of heat pump in €/MWh$_{el}$ |
| $C^{var,processing}$ | Variable operating costs of further processing steps (liquefaction/compression) of H$_2$ or CH$_4$ in €/MWh$_{fuel,\ LHV}$ |
| $C^{var,pv}$ | Variable operating costs of solar photovoltaics in €/MWh$_{el}$ |
| $C^{var,synthesis}$ | Variable operating costs of synthesis in €/MWh$_{el}$ |
| $C^{var,th.storage}$ | Variable operating costs of thermal storage in €/MWh$_{th}$ |
| $C^{var,wind_i}$ | Variable operating costs of wind turbines (type $i \in \{1,2\}$) in €/MWh$_{el}$ |
| $D^{(\cdot)}$ | Demand for fuel $(\cdot)$ as production target value for one year in MWh$_{fuel,\ LHV}$ |

## A. 3 Restrictions

In the following, the equations of the linear optimisation model describing the PtX process are explained.

### A 3.1 Installed capacities

For each technology, the hourly operation variable (e.g. power generation or power consumption) is restricted by the installed capacity of each system component (A-1)

$$
\begin{aligned}
0 &\leq p_t^{con,battery} \leq \beta^{battery} \cdot P^{battery} \quad \forall t \in T \\
0 &\leq p_t^{con,el.boiler} \leq P^{el.boiler} \quad \forall t \in T, \rho^{heat} = 1 \\
0 &\leq p_t^{con,electrolysis} \leq P^{electrolysis} \quad \forall t \in T \\
0 &\leq p_t^{con,processing} \leq P^{processing} \quad \forall t \in T, \rho^{processing} = 1 \\
0 &\leq p_t^{con,dac} \leq P^{dac} \quad \forall t \in T, \rho^{synthesis} = 1 \\
0 &\leq p_t^{electrolysis,LC+/-} \leq P^{electrolysis} \quad \forall t \in T \\
0 &\leq p_t^{gen,battery} \leq P^{battery} \quad \forall t \in T \\
0 &\leq p_t^{gen,pv} \leq AV_t^{pv} \cdot P^{pv} \quad \forall t \in T \\
0 &\leq p_t^{gen,wind_i} \leq AV_t^{wind_i} \cdot P^{wind_i} \quad \forall t \in T \quad \forall i \in \{1,2\} \\
0 &\leq q_t^{heat\ pump,th.storage} \leq P^{heat\ pump} \quad \forall t \in T, \rho^{heat} = 1 \\
0 &\leq fuel_t^{H_2,synthesis} \leq P^{synthesis} \quad \forall t \in T, \rho^{synthesis} = 1 \\
0 &\leq soc_t^{battery} \leq SOC^{battery} \quad \forall t \in T \\
0 &\leq soc_t^{CH_4} \leq SOC^{CH_4} \quad \forall t \in T, \rho^{synthesis} = 1, \rho^{CH_4\ storage} = 1 \\
0 &\leq soc_t^{H_2} \leq SOC^{H_2} \quad \forall t \in T \\
0 &\leq soc_t^{Q} \leq SOC^{Q} \quad \forall t \in T, \rho^{heat} = 1 \\
0 &\leq SOC^{battery} \leq \alpha^{battery} \cdot P^{battery}
\end{aligned}
\quad \text{(A-1)}
$$

## A 3.2 Electrolysis

As load change costs for electrolysis units are considered the following Eq. (A-2) describes the associated restriction.

$$p_{t+1}^{con,electrolysis} = p_t^{con,electrolysis} + p_t^{electrolysis,LC+} - p_t^{electrolysis,LC-} \quad \forall t \in T, t > 1 \qquad \text{(A-2)}$$

## A 3.3 Hydrogen storage

The storage level of the hydrogen storage increases by hydrogen production from the electrolysis and decreases by hydrogen used from the synthesis and from liquefaction or compression of hydrogen to get the final product (see Eq. (A-3)). If there is no hydrogen storage than the equation is just used as balance point with $soc_t^{H_2} = 0 \; \forall t$ and the loss factor $l^{H2\;storage} = 0$.

$$\begin{aligned} soc_{t+1}^{H_2} = soc_t^{H_2} + p_t^{con,electrolysis} \cdot \eta^{electrolysis} \cdot (1 - l^{H2\;storage}) \\ - \rho^{synthesis} \cdot fuel_t^{H_2,synthesis} - (1 - \rho^{synthesis}) \cdot fuel_t^{H_2,liquefy} \\ \forall t \in T, \rho^{H2\;storage} = 1 \end{aligned} \qquad \text{(A-3)}$$

## A 3.4 Direct air capture

The power consumption of the direct air capture depends on the operation of the synthesis as well as an hourly humidity factor, which influences the process efficiency as described in Eq. (A-4).

$$p_t^{con,dac} = fuel_t^{H_2,synthesis} \cdot \frac{\delta^{heat,dac}}{\gamma_t^{humidity}} \quad \forall t \in T, \rho^{synthesis} = 1 \qquad \text{(A-4)}$$

For global investigations, the hourly humidity factor was set equal to 1.

## A 3.5 Methane storage

The storage level of the methane storage increases by fuel production from the synthesis and decreases by suppling the final product possibly with liquefaction (see Eq. (A-5)). If there is no methane storage in the case of methanol or Fischer-Tropsch fuel production than the equation is just used as balance point with $soc_t^{CH_4} = 0 \; \forall t$.

$$\begin{aligned} soc_{t+1}^{CH_4} = soc_t^{CH_4} + fuel_t^{H_2,synthesis} \cdot \eta^{synthesis} - \rho^{processing} \cdot fuel_t^{CH_4,liquefy} \\ - (1 - \rho^{processing}) \cdot fuel_t^{(\cdot),final} \quad \forall t \in T, \rho^{synthesis} = 1, \rho^{CH4\;storage} = 1 \end{aligned} \qquad \text{(A-5)}$$

## A 3.6 Final processing

For end products with synthesis and liquefaction or compression the power consumption of the final processing step is described in Eq. (A-6).

$$\begin{aligned} p_t^{con,processing} = \eta^{processing} \cdot \rho^{synthesis} \cdot fuel_t^{CH_4,liquefy} \\ + \eta^{processing} \cdot (1 - \rho^{synthesis}) \cdot fuel_t^{H_2,liquefy} \quad , \rho^{processing} = 1 \end{aligned} \qquad \text{(A-6)}$$

## A 3.7 Final product balance

The total fuel production over the period of one year is summed up and has to equal a predefined value. Due to different used variables, corresponding to different production process steps, Eq. (A-7) encompasses different cases.

$$\begin{aligned} D^{(\cdot)} = \sum_{t \in T, t>1} (1 - \rho^{synthesis}) \cdot fuel_t^{H_2,liquefy} \\ + \rho^{synthesis} \cdot \left( \rho^{processing} \cdot fuel_t^{CH_4,liquefy} + (1 - \rho^{processing}) \cdot fuel_t^{(\cdot),final} \right) \end{aligned} \qquad \text{(A-7)}$$

## A 3.8 Battery storage

The state-of-charge of the battery storage is a function of charging and discharging (see Eq. (A-8)).

$$soc_{t+1}^{battery} = soc_t^{battery} \cdot (1 - l^{battery}) + p_t^{con,battery} \cdot \eta^{battery,con} - \frac{p_t^{gen,battery}}{\eta^{battery,gen}} \quad \forall t \in T, t > 1 \quad \text{(A-8)}$$

As load change costs for battery storage units are considered the following Eq. (A-9) and Eq. (A-10) are applied for charging and discharging of the battery.

$$p_{t+1}^{con,battery} = p_t^{con,battery} + p_t^{battery,con,LC+} - p_t^{battery,con,LC-} \quad \forall t \in T, t > 1 \quad \text{(A-9)}$$
$$p_{t+1}^{gen,battery} = p_t^{gen,battery} + p_t^{battery,gen,LC+} - p_t^{battery,gen,LC-} \quad \forall t \in T, t > 1 \quad \text{(A-10)}$$

## A 3.9 Power balance

In each time step, the power consumption has to equal the power generation in the systems as no external power source is considered (stand-alone system), see Eq. (A-11).

$$p_t^{con,battery} + \rho^{heat} \cdot p_t^{con,el.boiler} + \rho^{heat} \cdot \frac{q_t^{heat\,pump,th.storage}}{\eta^{heat\,pump}} + p_t^{con,electrolysis} + \rho^{heat}$$
$$\cdot \rho^{processing} \cdot p_t^{con,processing} + \rho^{heat} \cdot \rho^{synthesis} \cdot p_t^{con,dac} \quad \text{(A-11)}$$
$$= p_t^{gen,battery} + p_t^{gen,pv} + \sum_{i=1}^{2} p_t^{gen,wind_i} \quad \forall t \in T$$

## A 3.10 Heat related restrictions

In the following, all restricting related to heat demanding processes are explained. An electric boiler and a heat pump can be used as dispatchable heat sources. The heat from the electric boiler can be used by the electrolysis process directly or stored in the heat storage (s. Eq. (A-12)). The state of charge of the heat storage increased through heat production from electric boiler or heat pump and decreases through the demand for the DAC process as described in Eq. (A-13). For the high temperature electrolysis a stand-by heat demand is considered which is independent from the unit's dispatch. The heat demand of the electrolysis process can be supplied by the electric boiler and partly by waste heat of the synthesis process (see Eq. (A-14)). In case of PEM electrolysis the heat demand factor $\delta^{heat,electrolysis} = 0$. As the temperature level of the waste heat of the synthesis process is lower than the required temperature level of the SOEC only a share of the synthesis waste heat can be used for the electrolysis heat supply according to Eq. (A-15) and Eq. (A-16).

$$\eta^{el.boiler} \cdot p_t^{con,el.boiler} = q_t^{el.boiler,electrolysis} + q_t^{el.boiler,th.storage} \quad \forall t \in T, \rho^{heat} = 1 \quad \text{(A-12)}$$

$$soc_{t+1}^Q = soc_t^Q + q_t^{el.boiler,th.storage} + q_t^{heat\,pump,th.storage}$$
$$-\rho^{synthesis} \cdot fuel_t^{H_2,synthesis} \cdot \frac{\delta^{heat,dac}}{\gamma_t^{humidity}} \quad \forall t \in T, t > 1 \quad \text{(A-13)}$$

$$q_t^{el.boiler,electrolysis} + q_t^{synthesis,electrolysis} = (1 - \tau^{electrolysis}) \cdot \frac{\delta^{heat,electrolysis}}{\eta^{electrolysis}} \cdot p_t^{con,electrolysis}$$
$$+\tau^{electrolysis} \cdot \frac{\delta^{heat,electrolysis}}{\eta^{electrolysis}} \cdot P^{electrolysis} \quad \forall t \in T, \rho^{heat} = 1 \quad \text{(A-14)}$$

$$q_t^{synthesis,electrolysis} \leq \bar{\sigma}^{waste\,heat} \cdot p_t^{con,electrolysis} \cdot \frac{\delta^{heat,electrolysis}}{\eta^{electrolysis}} \quad \forall t \in T, \rho^{heat} = 1 \quad \text{(A-15)}$$
$$q_t^{synthesis,electrolysis} \leq \theta^{synthesis} \cdot fuel_t^{H_2,synthesis} \quad \forall t \in T, \rho^{heat} = 1, \rho^{synthesis} = 1 \quad \text{(A-16)}$$

## A. 4 Objective function

The objective function of the optimisation problem which has to be minimized consists of cost terms for the different system components. For all system components, equivalent annual investment costs, fixed as well as variable operating costs are considered. Furthermore, load change costs for the electrolysis and battery storage are also incorporated (see Eq. (A-17)).

$$\begin{aligned}
&(C^{inv,battery} + C^{fix,battery}) \cdot P^{battery} + \rho^{synthesis} \cdot (C^{inv,CH4\ storage} + C^{fix,CH4\ storage}) \\
&\cdot SOC^{CH_4} + \rho^{synthesis} \cdot (C^{inv,dac} + C^{fix,dac}) \cdot P^{dac} + \rho^{heat} \\
&\cdot (C^{inv,el.boiler} + C^{fix,el.boiler}) \cdot P^{el.boiler} + (C^{inv,electrolysis} + C^{fix,electrolysis}) \\
&\cdot P^{electrolysis} + (C^{inv,H2\ storage} + C^{fix,H2\ storage}) \cdot SOC^{H_2} + \rho^{heat} \\
&\cdot (C^{inv,heat\ pump} + C^{fix,heat\ pump}) \cdot P^{heat\ pump} + \rho^{synthesis} \\
&\cdot (C^{inv,synthesis} + C^{fix,synthesis}) \cdot P^{synthesis} + (C^{inv,pv} + C^{fix,pv}) \cdot P^{pv} + \rho^{heat} \\
&\cdot (C^{inv,th.storage} + C^{fix,th.storage}) \cdot SOC^Q \\
&+ \sum_{i=1}^{2} (C^{inv,wind_i} + C^{fix,wind_i}) \cdot P^{wind_i} + C^{var,battery} \cdot p_t^{gen,battery} \\
&+ \rho^{synthesis} \cdot C^{var,CH4\ storage} \cdot soc_t^{CH_4} + \rho^{synthesis} \cdot C^{var,dac} \cdot p_t^{con,dac} + \rho^{heat} \quad\quad \text{(A-17)} \\
&\cdot C^{var,el.boiler} \cdot p_t^{con,el.boiler} + C^{var,electrolysis} \cdot p_t^{con,electrolysis} \\
&+ C^{var,H2\ storage} \cdot soc_t^{H_2} + \rho^{heat} \cdot C^{var,heat\ pump} \cdot q_t^{heat\ pump,th.storage} \\
&+ \rho^{synthesis} \cdot C^{var,processing} \cdot fuel_t^{CH_4,liquefy} + \rho^{processing} \cdot C^{var,processing} \\
&\cdot fuel_t^{H_2,liquefy} + C^{var,pv} \cdot p_t^{gen,pv} + \rho^{synthesis} \cdot C^{var,synthesis} \\
&\cdot fuel_t^{H_2,synthesis} + \rho^{heat} \cdot C^{var,th.storage} \cdot soc_t^Q + \sum_{i=1}^{2} C^{var,wind_i} \cdot p_t^{gen,wind_i} \\
&+ C^{LC,battery} \\
&\cdot (p_t^{battery,con,LC+} + p_t^{battery,con,LC-} + p_t^{battery,gen,LC+} + p_t^{battery,gen,LC-}) \\
&+ C^{LC,electrolysis} \cdot (p_t^{electrolysis,LC+} + p_t^{electrolysis,LC-})
\end{aligned}$$

# B Appendix Further techno-economic assumptions

## B. 1   Further information on the consideration of desalination plants

*Table B 1: Techno-economic assumptions for calculation of the impact of the costs and the energy consumption of the desalination plant*

| CAPEX | OPEX | Life time | Electricity consumption | Water consumption (PEM electrolysis) |
|---|---|---|---|---|
| 2.23 €/(m³ × yr) [84] | 4 % CAPEX [85] | 30 yrs [85] | 4 kWh/m³ [22,85] | 0.33 m³/MWh$_{H2,\ LHV}$ [64] |

For easier calculation, we assumed that both electrolyser technologies have the same water consumption of process water. All calculations are carried out for the use case of the production of 1 TWh hydrogen per year ("best case fuel": Hydrogen) or 1.3 TWh ("worst case fuel": FT fuels). These 2 fuels represent the lower and the upper boundary of the influence of desalination, due to the best and the worst efficiency chain during their production. The desalination plant needs a capacity of $1,000,000\ MWh_{H2} \cdot 0.33\ \frac{m^3}{MWh_{H2}} = 330,000\ \frac{m^3}{yr}$ ($437,000\ \frac{m^3}{yr}$ for FT fuel). This results in investment costs of 735,900 € for hydrogen and 974,130 € for FT fuel.

#### Calculation of the additional costs

Calculation of the equivalent annual costs of the desalination plant.

$$A = CAPEX \cdot (1+i)^{lt} \cdot \left(\frac{i}{(1+i)^{lt} - 1}\right) \quad\quad \text{(B-1)}$$

A – equivalent annual costs of the investment

lt – lifetime of the component

i – interest rate of the investment

The equivalent annual cost of the desalination plant calculates to 65,368 € and 86,529 € for FT fuels respectively. This leads (with OPEX) to annual costs of 94,804 € for hydrogen and 125,494 € for FT fuels.

Normed to 1 MWh hydrogen this results in approximately 0.095 €/MWh$_{LHV}$ hydrogen and 0.125 €/MWh$_{LHV}$ for FT fuels.

**Calculation of the additional energy consumption**

The additional electricity demand can be calculated from the electricity consumption of 4 kWh/m³ and the process water demand of 330,000 m³ (437,000 m³ for FT fuels). This leads to 1,320 MWh$_{el}$ additional electricity demand for hydrogen and 1.748 MWh$_{el}$ for FT fuels. The overall electricity demand for the production of 1 TWh hydrogen sums up to 1,408,450 MWh$_{el}$ for PEM or 1,136,363 MWh$_{el}$ for SOEC, respectively. In the case of the production of FT fuels the electricity demand is 1,845,937 MWh$_{el}$ for PEM or 1,489,335 MWh$_{el}$ for SOEC. So even in the worst case (FT fuels with PEM electrolysis) the electricity demand of the desalination plant represents only 0.117 % of the electricity demand of the fuel production of 1 TWh$_{LHV}$ FT fuels.

Since the technological development until 2050 is subject to so many uncertainties and beyond that the influence of the desalination plant on the energy demand and production cost is marginal the desalination plant was not included in the model. Rather the influence was considered by reducing the electrolyser efficiencies and adding the additional costs to the final modelled production costs for every pathway.

## B. 2   Additional specifications of the site simulation

The following tables (Table B 2 until Table B 4) contain further information on specifications used during the optimisation.

Table B 2: Lifetimes of the different PtX production components that were utilised in this study

| Technology | Lifetime |
|---|---|
| DAC unit | 30 yrs |
| Photovoltaic power plant | 25 yrs |
| Battery storage | 15 yrs |
| Every other component of the PtX production facility | 20 yrs |

Table B 3: Battery storage configuration used in this study

| Parameter | Value |
|---|---|
| Efficiency (charging) | 93.81 % |
| Efficiency (discharging) | 93.81 % |
| Ratio charging to storage capacity | 4 MW$_{el}$/MWh$_{el}$ |
| Ratio charging to discharging | 1 MW$_{el}$/MW$_{el}$ |
| Storage losses | 0.001 %/h |

Table B 4: Matrix of the possible installation of hydrogen and methane storages for every production pathway

| Production pathway | Installation of a hydrogen storage possible? | Installation of a methane storage possible? |
|---|---|---|
| FT fuels (PEM) | Yes | No |
| Methanol (PEM) | Yes | No |
| Methane gaseous (PEM) | Yes | Yes |
| Methane liquid (PEM) | Yes | Yes |
| Ammonia (PEM) | Yes | No |
| Hydrogen gaseous (PEM) | No | No |
| Hydrogen liquid (PEM) | Yes | No |
| FT fuels (SOEC) | No | No |
| Methanol (SOEC) | No | No |
| Methane gaseous (SOEC) | No | Yes |
| Methane liquid (SOEC) | No | Yes |
| Ammonia (SOEC) | No | No |
| Hydrogen gaseous (SOEC) | No | No |
| Hydrogen liquid (SOEC) | Yes | No |

## B. 3  Additional specifications of the LCOE calculation

For the potential area identification the LCOE for the RE sources was an important criterion. For PV plants we derived the capacity factor directly from the Global Solar Atlas [37]. For wind power plants we used a reference performance curve (Table B 5) and used the internal wind model [36] to calculate the capacity factor. In addition to the use of a shading model and a power-curve smoother we considered a reduction factor of 0.95 to take technical unavailability due to maintenance and repair into account.

Table B 5: Performance curve table for reference wind power plant

| Wind speed | Capacity factor | Wind speed | Capacity factor |
|---|---|---|---|
| 1 m/s | 0.00490 | 13 m/s | 0.86602 |
| 2 m/s | 0.02092 | 14 m/s | 0.87554 |
| 3 m/s | 0.05713 | 15 m/s | 0.85961 |
| 4 m/s | 0.10092 | 16 m/s | 0.82067 |
| 5 m/s | 0.16858 | 17 m/s | 0.76376 |
| 6 m/s | 0.25824 | 18 m/s | 0.69386 |
| 7 m/s | 0.36838 | 19 m/s | 0.61803 |
| 8 m/s | 0.48735 | 20 m/s | 0.54160 |
| 9 m/s | 0.59921 | 21 m/s | 0.47054 |
| 10 m/s | 0.69732 | 22 m/s | 0.40382 |
| 11 m/s | 0.77451 | 23 m/s | 0.34369 |
| 12 m/s | 0.83184 | 24 m/s | 0.29020 |

We finally calculated the LCOE with the help of these capacity factors and the techno-economic assumptions for PV and wind power plants used in this study (Table B 2, Table 7) with a WACC of 8 % for each country considered.

## B. 4  Example for the selection of techno-economical parameters out of the literature research

Table B 6: Extract of the selection of techno-economical parameters from different studies

| Parameter | Lower boundary (examples) | Upper boundary (examples) | Used in this study |
|---|---|---|---|
| CAPEX DAC | 195 €/($t_{CO2}$ × yr) [86] | 1033 €/($t_{CO2}$ × yr) [15] | 450 €/($t_{CO2}$ × yr) |
| DAC heat consumption | 1,102 kWh$_{th}$/$t_{CO2}$ [70] | 1,600 kWh$_{th}$/$t_{CO2}$ [87] | 1,312.2 kWh$_{th}$/$t_{CO2}$ |
| DAC electricity consumption | 182 kWh$_{el}$/$t_{CO2}$ [70] | 400 kWh$_{el}$/$t_{CO2}$ [87] | 255.15 kWh$_{el}$/$t_{CO2}$ |
| Efficiency PEM (overall) | 63.5 % [15] | 84 % [62] | 71 % |
| CAPEX PEM | 400 €/kW$_{el}$ [88] | 793 €/kW$_{el}$ [15] | 470 €/kW$_{el}$ |